\documentclass[12pt,preprint]{aastex}
\usepackage{amssymb,amsmath,mathrsfs,color}
\usepackage{bm}
\usepackage{multirow}
\usepackage{hyperref}

\newcommand{\niceurl}[1]{\href{#1}{\textsl{#1}}}

\graphicspath{{figs/}{figs2/}{figs4/}}

\newcommand{\arxiv}[1]{\href{http://arxiv.org/abs/#1}{arXiv:#1}}

\newcommand{\foreign}[1]{\emph{#1}}
\newcommand{\etal}{\foreign{et\,al.}}

\newcommand{\figref}[1]{\figurename~\ref{#1}}
\newcommand{\Figref}[1]{\figref{#1}}
\newcommand{\tabref}[1]{\tablename~\ref{#1}}
\newcommand{\Tableref}[1]{\tabref{#1}}

\newcommand{\thou}{,\!000}

\renewcommand{\micron}{$\mu$m}
\slugcomment{Submitted to The Astronomical Journal}

\newcommand{\thetractor}{\textsl{the Tractor}}
\newcommand{\Thetractor}{\textsl{The Tractor}}
\newcommand{\tractor}{\textsl{Tractor}}


\begin{document}

\title{WISE photometry for 400 million SDSS sources}
\author{
  Dustin~Lang\altaffilmark{1,2,3},
  David~W.~Hogg\altaffilmark{4,5},
  David~J.~Schlegel\altaffilmark{6}}
\shorttitle{WISE photometry of SDSS sources}
\shortauthors{Lang, Hogg, Schlegel}
\altaffiltext{1}%
{McWilliams Center for Cosmology,
  Department of Physics, Carnegie Mellon University,
  5000 Forbes Ave, Pittsburgh, PA, 15213, USA}
\altaffiltext{2}%
{Department of Physics \& Astronomy,
  University of Waterloo,
  200 University Avenue West,
  Waterloo, ON, N2L 3G1, Canada}
\altaffiltext{3}%
{To whom correspondence should be addressed: dstn@cmu.edu}
\altaffiltext{4}%
{Center for Cosmology and Particle Physics,
 Department of Physics, New York University,
 4 Washington Place, New York, NY, 10003, USA}
\altaffiltext{5}%
{Max-Planck-Institut f\"ur Astronomie,
 K\"onigstuhl 17, D-69117, Heidelberg, Germany}
\altaffiltext{6}%
{Lawrence Berkeley National Laboratory,
 1 Cyclotron Road, Berkeley, CA, 94720, USA}
\date{}

\begin{abstract}
We present photometry of images from the Wide-Field Infrared Survey
Explorer (WISE; \citealt{wright}) of over 400 million sources detected
by the Sloan Digital Sky Survey (SDSS; \citealt{york}).
%
%
We use a ``forced photometry'' technique, using measured SDSS source
positions, star--galaxy separation and galaxy profiles to define the
sources whose fluxes are to be measured in the WISE images.
We perform photometry with \Thetractor\ image modeling code, working
on our ``unWISE'' coaddds and taking account of the WISE point-spread
function and a noise model.
The result is a measurement of the flux of each SDSS source in each
WISE band.  Many sources have little flux in the WISE bands, so often
the measurements we report are consistent with zero.  However, for
many sources we get three- or four-sigma measurements; these sources
would not be reported by the WISE pipeline and will not appear in the
WISE catalog, yet they can be highly informative for some scientific
questions.  In addition, these small-signal measurements can be used
in \emph{stacking} analyses at catalog level.  The forced photometry
approach has the advantage that we measure a \emph{consistent} set of
sources between SDSS and WISE, taking advantage of the resolution and
depth of the SDSS images to interpret the WISE images; objects that
are resolved in SDSS but blended together in WISE still have accurate
measurements in our photometry.
Our results, and the code used to produce them, are publicly available
at \niceurl{http://unwise.me}.
\end{abstract}

\section{Introduction}

In astronomical survey projects, it is common practice to produce a
\emph{catalog} of detected and measured sources, using only data from
the survey itself.  This approach has the benefit that the survey can
be thought of as an independent experiment, but the disadvantage that
it ignores the huge amount of information we already have about
astronomical sources measured in previous surveys covering the same
part of the sky.  While new surveys typically bring some new
capability that previous surveys lacked (eg, depth, resolution, or
wavelength coverage), it is seldom the case that a new survey
surpasses all previous data in all regards; there is usually some
complementary information that could be of value.

This approach of compiling ``independent'' catalogs has two
shortcomings, in particular when it comes to comparing the new survey
with existing surveys.  First, when the new survey has lower
resolution, there will be some nearby sets of sources that are blended
together (detected and measured as a single source) in the new
catalog, but resolved in existing data.  Second, when the new survey
has lower sensitivity (at least to some types of sources), sources
known from previous surveys will not be detected in the new survey and
will not appear in its catalog.  When investigators attempt to
cross-match the new catalog with existing catalogs (usually via
astrometric cross-matching), the first problem (blended sources)
typically results in either failed matches (because the blended source
has a different centroid), or very strange inferred properties (for
example, bizarre colors because the new survey matches the sum of a
set of sources to a single source in the existing survey; or
unexpected non-matching sources).  The second problem (non-detections)
means that fewer sources are available to cross-match; a catalog
cross-match is limited by the weaknesses of both catalogs.

In contrast, in this paper we perform ``forced photometry'' of a new
survey (WISE) given a great deal of knowledge from an existing survey
(SDSS).  While WISE has comparable \emph{depth} to SDSS for many
sources, its resolution is significantly lower.  We therefore get
significant benefit from using SDSS detections to decide \emph{where
  to look} in the WISE data.

The Wide-Field Infrared Survey Explorer (WISE; \citealt{wright})
measured the full sky in four mid-infrared bands centered on 3.4
\micron, 4.6 \micron, 12 \micron, and 22 \micron, known as W1 through
W4.  During its primary mission, it scanned the full sky in all four
bands.  After its solid hydrogen cryogen ran out and W4 became
unusable, it continued another half-sky scan in W1, W2, and W3.
During the ``NEOWISE post-cryo'' continuation (\citealt{mainzer}), it
continued to scan another half-sky in W1 and W2.  Over 99\% of the sky
has 11 or more exposures in W3 and W4, and 23 or more exposures in W1
and W2.  Median coverage is 33 exposures in W1 and W2, 24 in W3, and
16 in W4.
In December 2013, WISE was reactivated and is expected to
complete several more full scans of the sky in W1 and W2 (\citealt{mainzer2014}).

The WISE team have made a series of high-quality data releases, the
most recent of which is the AllWISE Data Release.\footnote{Explanatory
  Supplement to the AllWISE Data Release Products, 
  \niceurl{http://wise2.ipac.caltech.edu/docs/release/allwise/expsup/}}
The AllWISE Release includes a source catalog of nearly 750 million
sources, a database of photometry in the individual frames at each
source position, and ``Atlas Images'': coadded matched-filtered
images.
The AllWISE Atlas Images were intentionally convolved by the
point-spread function (PSF), making it challenging to use them for
forced photometry.  Instead, we use the ``unWISE'' coadds from
\citet{unwise}, which preserve the resolution of the original WISE
images.

The Sloan Digital Sky Survey (SDSS; \citealt{york}) imaged over
$14\thou$ square degrees of sky in five bands ($u$, $g$, $r$, $i$, $z$), detecting and
measuring over $400$ million sources.  We use the imaging catalogs
from SDSS-III Data Release 10 (\citealt{sdss3, dr10}).  These catalogs
contain the outputs of the \emph{Photo} pipeline (\citealt{lupton}),
and include star/galaxy separation and galaxy shape measurements using
either exponential, de Vaucouleurs, or composite (sum of exponential
and de Vaucouleurs) profiles.

The combination of data from SDSS and WISE has proven to be very
powerful for a variety of studies.  \citet{yan} give a survey of the
properties of extragalactic sources, showing that SDSS--WISE colors and
morphology can be used to select type-2 dust-obscured quasars and
ultra-luminous infrared galaxies at redshift $\sim2$.
The SDSS-III BOSS survey (\citealt{boss}) includes quasars targeted using
SDSS color cuts and WISE detection in the W1, W2, and W3 bands, to select $z>2$ quasars.
%

The work described here was motivated by the need to select targets
for the SDSS-III SEQUELS and SDSS-IV eBOSS programs.  Myers \etal\ (in
prep.) describe the use of our results to select quasars, while
Prakash \etal\ (in prep.) describe the selection of luminous red
galaxy (LRG) targets.  The LRG targets are fairly bright in WISE, so a
catalog match produces satisfactory results.  However, due to the
lower resolution of the WISE images, nearby sources that are resolved
in SDSS may be blended in WISE, resulting either in missed astrometric
matches (because the WISE centroid is shifted), or incorrect colors
(because the WISE catalog source includes flux from multiple SDSS
sources).  Using our results improves this situation, since we
photometer a consistent set of sources.  For the quasar targets, the
often few-sigma flux measurements we make are of considerable utility.
In the redshift range of interest, the quasar and stellar loci are
significantly separated in SDSS--WISE colors, so even a noisy
measurement of the WISE flux can effectively eliminate stellar
contamination.


In similar work, the ``extreme deconvolution'' quasar target selection
and redshift-estimation method (XDQSOz; \citealt{bovy}) makes
effective use of forced photometry of GALEX UV (\citealt{martin}) and
UKIDSS near-IR (\citealt{lawrence}) images, based on SDSS source
positions.  While often low-signal-to-noise, these measurements
nevertheless can be very effective in eliminating degeneracies in
quasar classification and redshift determination.  Indeed, the XDQSOz
method has been extended to incorporate the measurements we present
here by DiPompeo \etal\ (in prep).

\section{Method}

We use \thetractor\ code (Lang \etal, in prep.) in ``forced
photometry'' mode.  In general, \thetractor\ optimizes or samples from
a full generative model that includes parameters of the image
calibration and all the parameters of the sources in the images
(positions, shapes, and fluxes).  In forced photometry mode, the image
calibration parameters are frozen (held fixed), as are all properties of the
sources except for their fluxes in the bands of interest.  In this
case, the photometry task becomes linear: We know what each source
should look like in the WISE images, and we must compute the linear
sum of the sources that best matches the observed image.

\begin{figure}
  \begin{center}
    \includegraphics[width=0.8\textwidth]{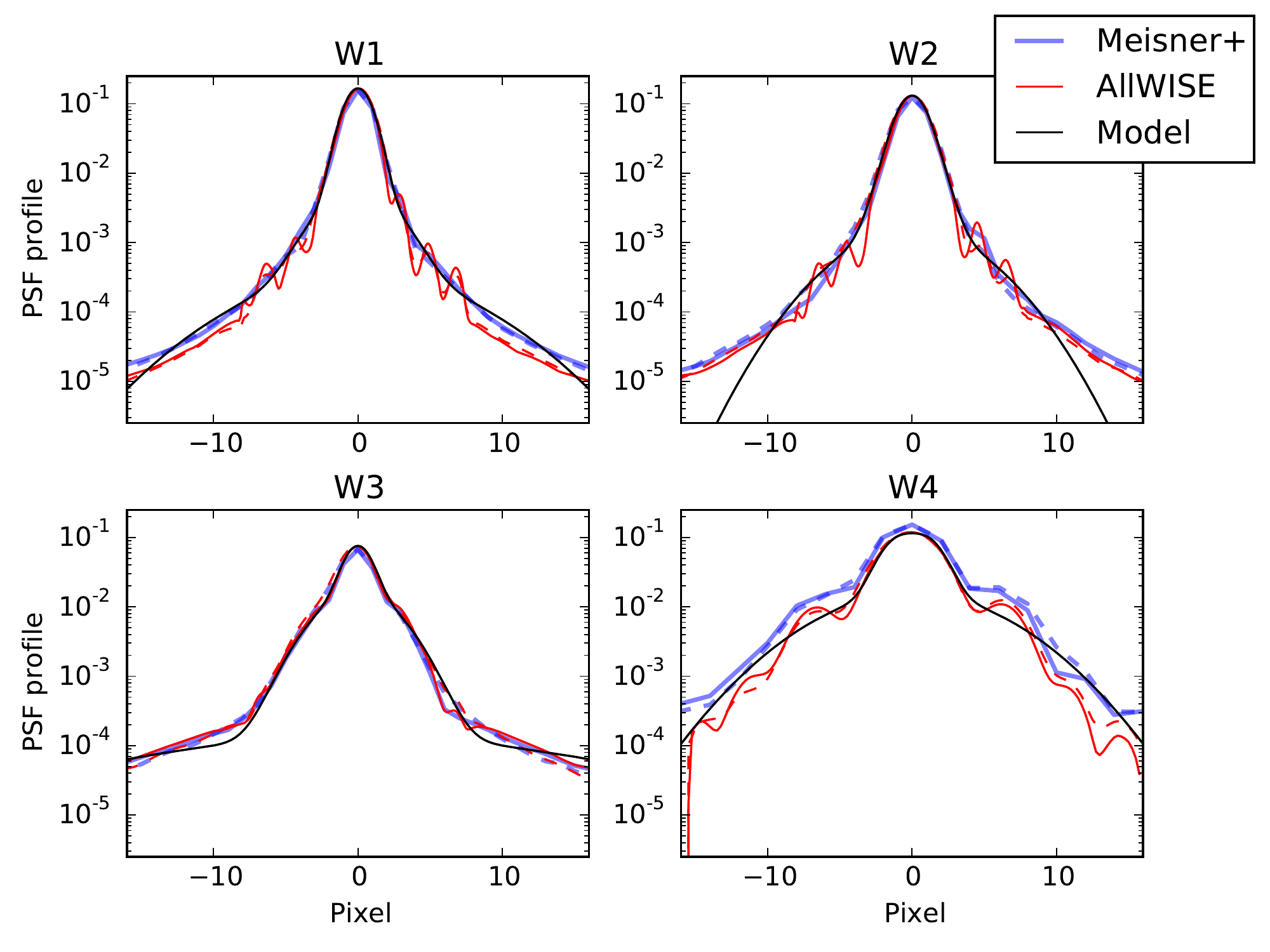}
  \end{center}
  \caption{%
    PSF models from the AllWISE Release, from \citealt{meisner}, and
    our three-component, zero-mean, isotropic, concentric Gaussian
    fits.  Note the log scale.  The AllWISE and \citealt{meisner}
    plots are horizontal (solid) and vertical (dashed) slices through
    the PSF center.  Our mixture-of-Gaussian models capture the PSF
    cores---roughly three orders of magnitude---quite effectively, but
    degrade somewhat at large radii and low flux levels.
    \label{fig:psf}}
\end{figure}

The image calibration parameters include the astrometric calibration,
described by a World Coordinate System (WCS); the photometric
calibration, described by a zeropoint; a point-spread function model;
a noise model (per-pixel error estimates); and a ``sky'' or background
model.  We are photometering the ``unWISE'' coadds, which are tiles of
roughly $1.5^\circ \times 1.5^\circ$ in extent.  The tiles use a
gnomonic projection (tangent plane; WCS code ``TAN''), are
sky-subtracted, and have a photometric zeropoint of $22.5$ in the Vega
system.  In turn, these coadds use the ``level 1b'' calibrated
individual exposures from the WISE All-Sky Data Release.  
We use the WISE PSF models from the WISE All-Sky
Release,\footnote{Available at
  \niceurl{http://wise2.ipac.caltech.edu/docs/release/allsky/expsup/sec4\_4c.html\#psf}}
averaged over the focal plane and approximated by a mixture of three
isotropic concentric Gaussian components.
We have been impressed by the quality of the WISE PSF models from
\citet{meisner},\footnote{Available at
  \niceurl{https://github.com/ameisner/WISE/}} but have opted to use
the WISE team's models here for consistency.
\Figref{fig:psf} shows the AllWISE PSF models and our Gaussian
approximations.  \Tableref{tab:psf} lists the parameters of our PSF
models.

\begin{deluxetable}{c@{\hspace{2em}}r@{.}lr@{.}l@{\hspace{2em}}r@{.}lr@{.}l@{\hspace{2em}}r@{.}lr@{.}l@{\hspace{2em}}r@{.}lr@{.}l}
  \tabletypesize{\small}
  \tablewidth{0pt}

  \tablecaption{Mixture-of-Gaussian PSF fit parameters.  The rows list
    the three Gaussian components we use to represent the PSF.
    ``Amp'' indicates the amplitude or ``weight'' of the component,
    and ``Std'' is the standard devation of the Gaussian.  Notice that
    for W4, one of the components is negative; while perhaps
    surprising, the resulting PSF is still positive everywhere so this
    is not an issue.  Also notice that the sums of amplitudes are not
    strictly unity.
    \label{tab:psf}}

  \tablehead{
    \textbf{Component} &
    \multicolumn{4}{c}{\textbf{W1\hspace*{1em}}} &
    \multicolumn{4}{c}{\textbf{W2\hspace*{1em}}} &
    \multicolumn{4}{c}{\textbf{W3\hspace*{1em}}} &
    \multicolumn{4}{c}{\textbf{W4\hspace*{1em}}}
    \\
    &
    \multicolumn{2}{c}{Amp} &
    \multicolumn{2}{c}{Std} &
    \multicolumn{2}{c}{Amp} &
    \multicolumn{2}{c}{Std} &
    \multicolumn{2}{c}{Amp} &
    \multicolumn{2}{c}{Std} &
    \multicolumn{2}{c}{Amp} &
    \multicolumn{2}{c}{Std}
   }
   \startdata
   1 & $0$ & $7610$ & $0$ & $8664$ & $0$ & $5007$ & $0$ & $8972$ & $0$ & $2752$ & $0$ & $8555$ & $-0$ & $2929$ & $1$ & $1001$ \\
   2 & $0$ & $1538$ & $2$ & $1499$ & $0$ & $3410$ & $1$ & $3302$ & $0$ & $5537$ & $2$ & $3620$ & $0$ & $7035$ & $1$ & $3284$ \\
   3 & $0$ & $0723$ & $5$ & $8201$ & $0$ & $1344$ & $3$ & $7869$ & $0$ & $1461$ & $13$ & $3751$ & $0$ & $6207$ & $5$ & $0750$
   \enddata
\end{deluxetable}

\begin{figure}
\begin{center}
\begin{tabular}{ccc}
  SDSS $r$ image &
  WISE W1 image &
  WISE W1 image \\
  \includegraphics[width=0.3\textwidth]{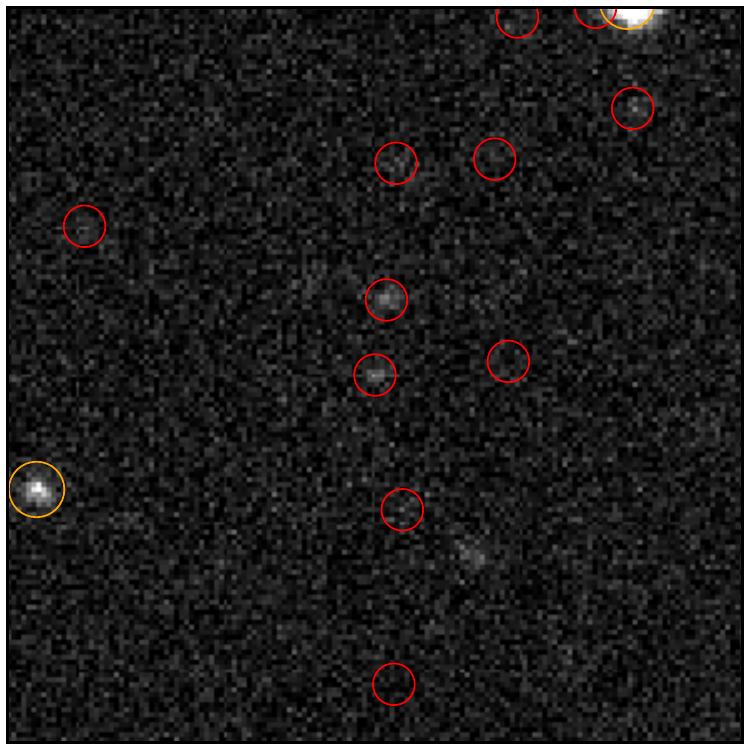} &
  \includegraphics[width=0.3\textwidth]{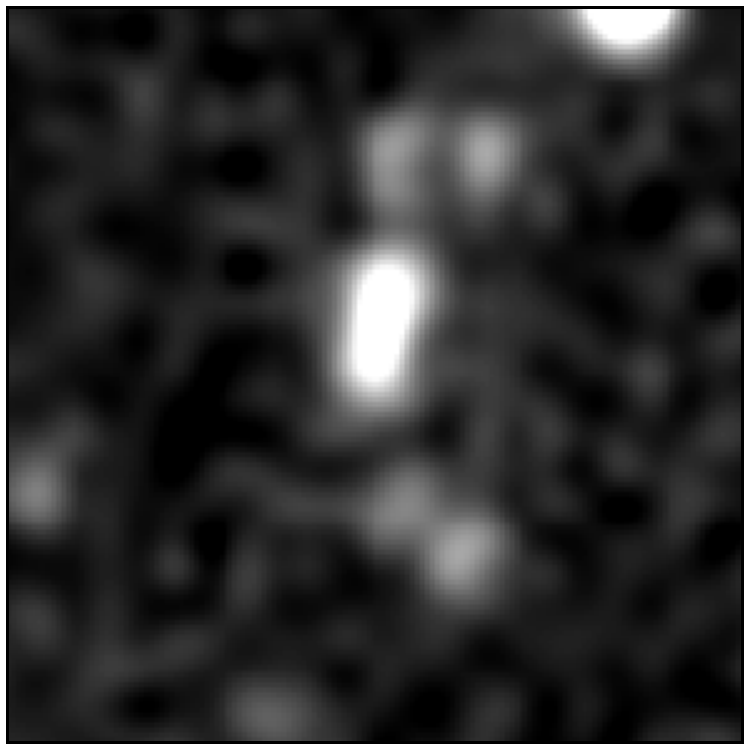} &
  \includegraphics[width=0.3\textwidth]{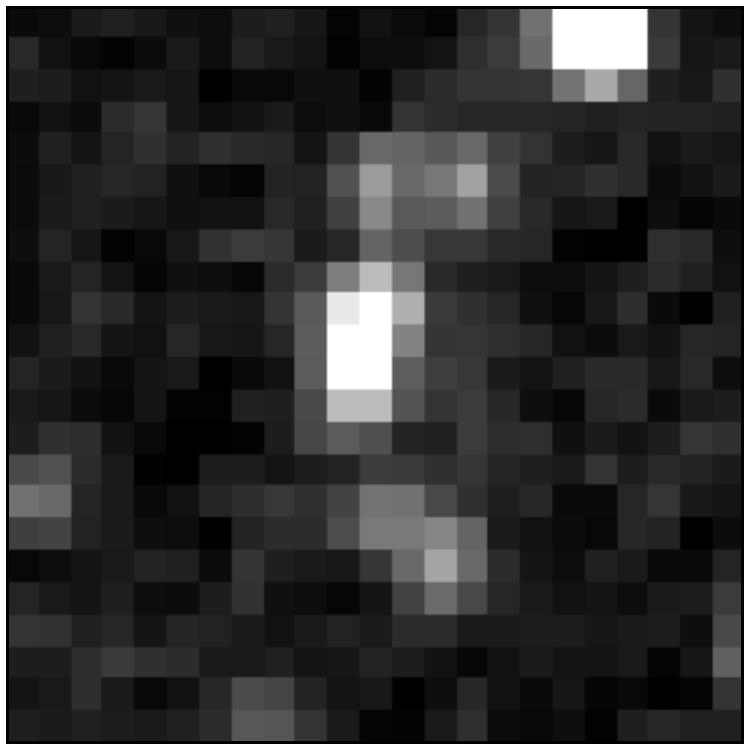} \\
  WISE W1 image &
  WISE W1 model &
  WISE W1 model + noise \\
  \includegraphics[width=0.3\textwidth]{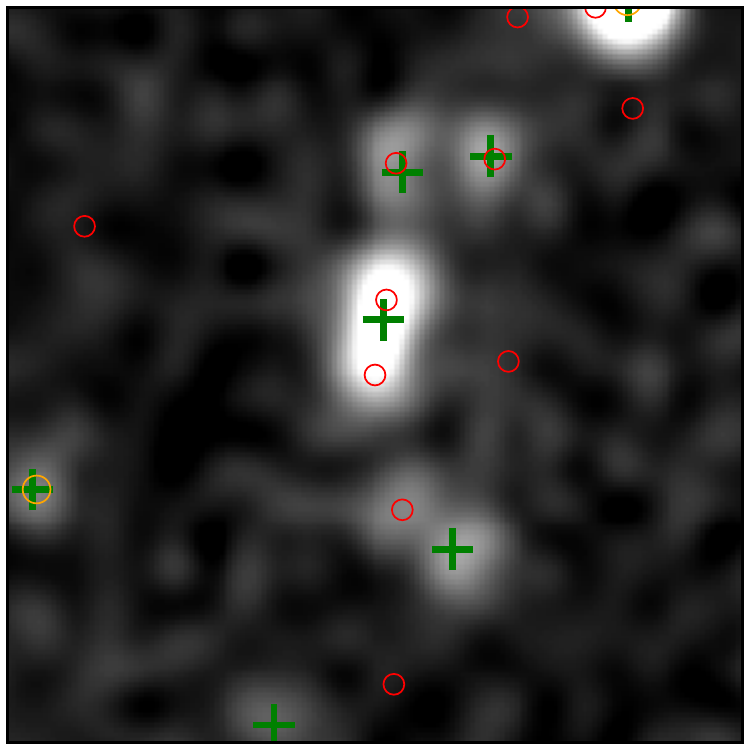} &
  \includegraphics[width=0.3\textwidth]{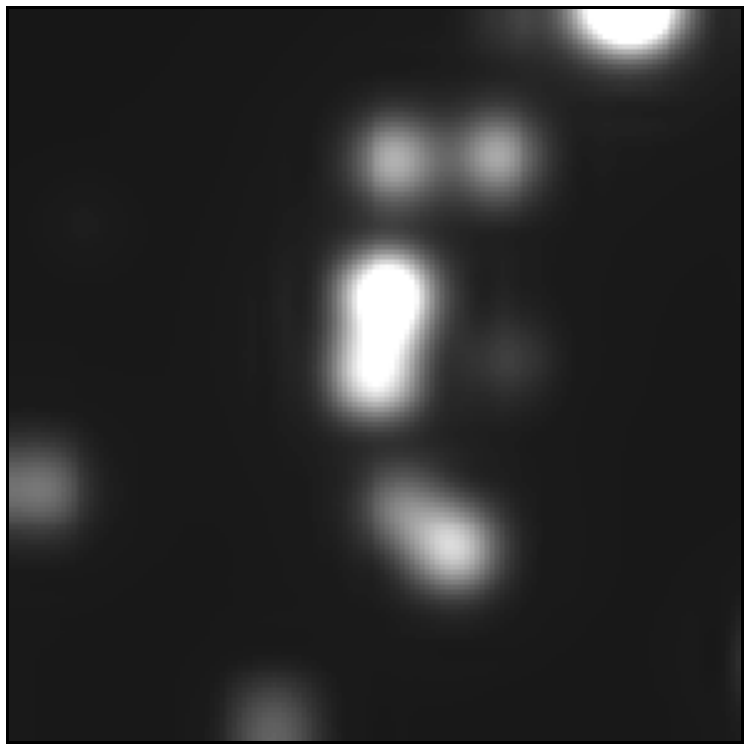} &
  \includegraphics[width=0.3\textwidth]{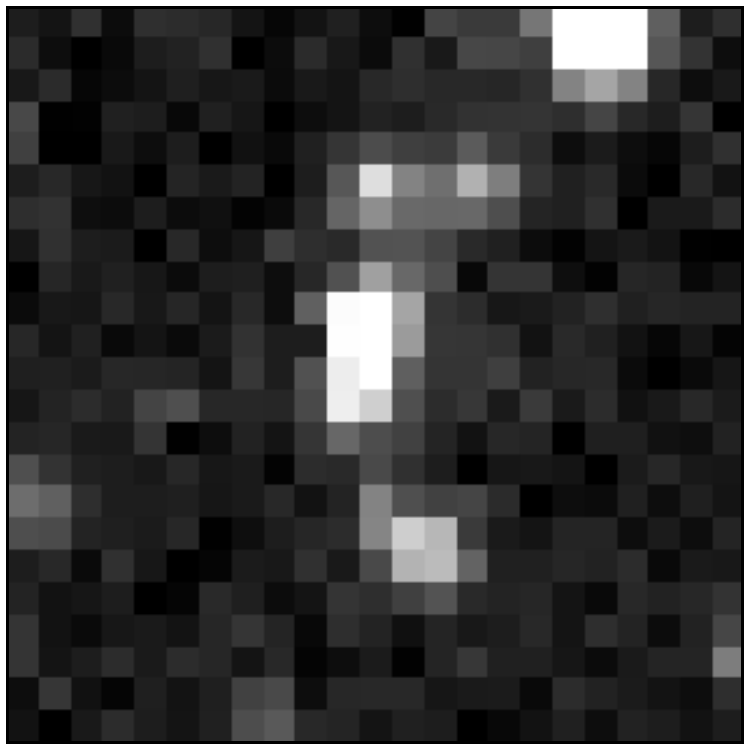}
\end{tabular}
\end{center}
\caption{Example forced photometry results.  \textbf{Top-left}: SDSS
  $r$-band image, with point sources (small red) and galaxies (large
  orange) marked.  (Some of these sources will have been detected in
  the other SDSS bands.)  These are the sources that will be
  photometered, along with sources from the WISE catalog that are
  detected only in WISE.  \textbf{Top-middle:} WISE W1 image, smoothed
  by resampling.  \textbf{Top-right:} WISE W1 image at original
  (native) resolution.  \textbf{Bottom-left:} WISE W1 image, with WISE
  catalog detections (green cross) and SDSS sources marked.  Notice
  that the central source is detected as a single source in WISE, but
  resolved into two sources by SDSS.  Also notice a number of
  WISE-only and SDSS-only detections.  \textbf{Bottom-middle:} Forced
  photometry model image.  This is a weighted sum of WISE PSF models
  (convolved by the galaxy profile for the one galaxy in this field)
  at the positions of SDSS and WISE-only sources, with weights chosen
  to minimize the chi-squared residuals from the WISE W1 image.
  \textbf{Bottom-right:} The model image at native resolution, plus
  per-pixel noise equal to that in the real image.  The model is a
  good approximation to the real image, given the observational noise.
  \label{fig:demo}}
\end{figure}

\begin{figure}[htb]
\begin{center}
\includegraphics[width=\textwidth]{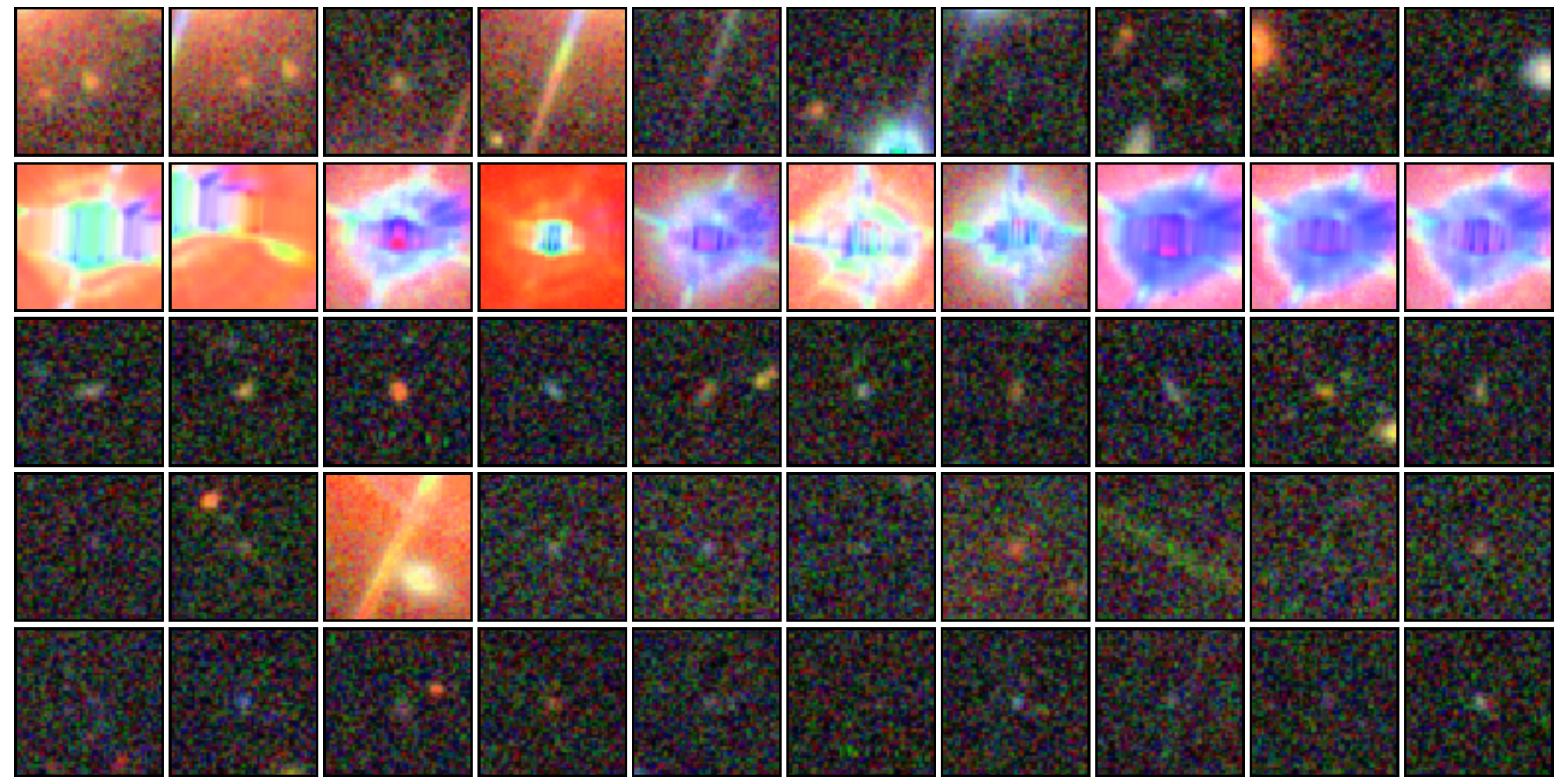}
\end{center}
\caption{Examples of SDSS sources that we treat as point sources for
  photometry.  Each image is a $50 \times 50$ cutout in SDSS $irg$
  bands.
  \newline\hspace*{1em}
  \textbf{Top row:} effective radius measured at a signal-to-noise less than 3.  
  \newline\hspace*{1em}
  \textbf{Second row:} sources with large flux (stars).
  \newline\hspace*{1em}
  \textbf{Third row:} axis ratio has a reported error of zero.
  \newline\hspace*{1em}
  \textbf{Fourth row:} effective radius reported is the largest allowed.
  \newline\hspace*{1em}
  \textbf{Bottom row:} effective radius signal-to-noise is larger than expected given the measured flux of the object.
  \label{fig:ptsrc}
}
\end{figure}

For an example of the forced photometry results, see \figref{fig:demo}.

We photometer sources in the SDSS DR10 imaging catalog.\footnote{We
  have also photometered what will become the DR13 catalog, and will
  release these results when DR13 is released.  For our purposes,
  these data releases differ only in their choice of which field to
  call ``primary''; the precise set of objects photometered will
  differ in some parts of the sky.}
We select ``survey primary'' sources (those detected in the ``best''
imaging scan covering each part of the sky) from the \emph{photoObj}
catalog files.  We drop the ``parent'' sources in blends, keeping the
deblended children.  We use the $r$-band galaxy shape measurements.
We find that many faint sources classified as galaxies have poorly
constrained galaxy shape measurements.  We treat these as point
sources rather than galaxies in our photometry.  Specifically, we
treat as a point source any galaxy whose effective radius is measured
at a signal-to-noise of less than 3; or with stated axis ratio error
of zero; or with the maximum effective radius considered by the
\emph{Photo} software; or with stated effective radius signal-to-noise
significantly greater than expected given its flux; or with magnitude
$r < 12.5$ (bright stars whose PSF wings are mistakenly identified as
galaxies).  Examples of sources we treat as point sources are shown in
\figref{fig:ptsrc}.

For each WISE tile, we keep SDSS sources that are within the tile plus
a margin of $20$ WISE pixels (55 arcseconds).  We also include in the
fitting sources that are detected in the AllWISE Release catalog but
not SDSS; we keep WISE catalog sources that have no SDSS match within
$4$ arcseconds.

\Thetractor\ code proceeds by rendering the galaxy or point source
models convolved by the image PSF model.  The galaxy profiles are
represented as mixtures of Gaussians, as described in
\citet{profiles}, and we fit a mixture of Gaussians to the PSF model.
The convolution is then analytic and rendering PSF-convolved galaxy
profiles becomes a matter of evaluating a large number of Gaussians.
In principle these Gaussian profiles have infinite extent, but we clip
them when the surface brightness drops below approximately $10\%$ of
the per-pixel noise.  Once the profile of each source has been
rendered, forced photometry requires performing a linear least-squares
fit for source fluxes such that their sum is closest to the actual
image pixels, with respect to the noise model.
This least-squares problem is very sparse (most sources touch only
dozens of pixels); \thetractor\ uses \emph{Ceres Solver}
(\citealt{agarwal}), which handles this case well, as its optimization
engine for forced photometry.

We photometer each WISE tile and each WISE band separately.  Since the
WISE tiles overlap slightly, this means we photometer some SDSS
sources in multiple tiles.  We resolve these multiple measurements
after processing all tiles, keeping only the measurement closest to
the center of its tile.  We then write out files that are row-by-row
parallel to the SDSS \emph{photoObj} input files.  The contents of our
catalogs are described in Appendix \ref{sec:catalog}.

\section{Results}

We photometered a total of $7,989$ WISE tiles, covering roughly
$14,900$ square degrees and containing roughly $469$ million SDSS
sources.  The AllWISE catalog contains roughly $240$ million sources
in the same area.  Of the $469$ million sources photometered, we
treated $430$ million as point sources and $34$ million as galaxies.
Photometry took roughly $1500$ CPU-hours total.


\subsection{Comparison to WISE catalog}

\begin{figure}
\begin{tabular}{@{}c@{}c@{}c@{}}
All Matches & Galaxies & Stars \\
\includegraphics[width=0.33\textwidth]{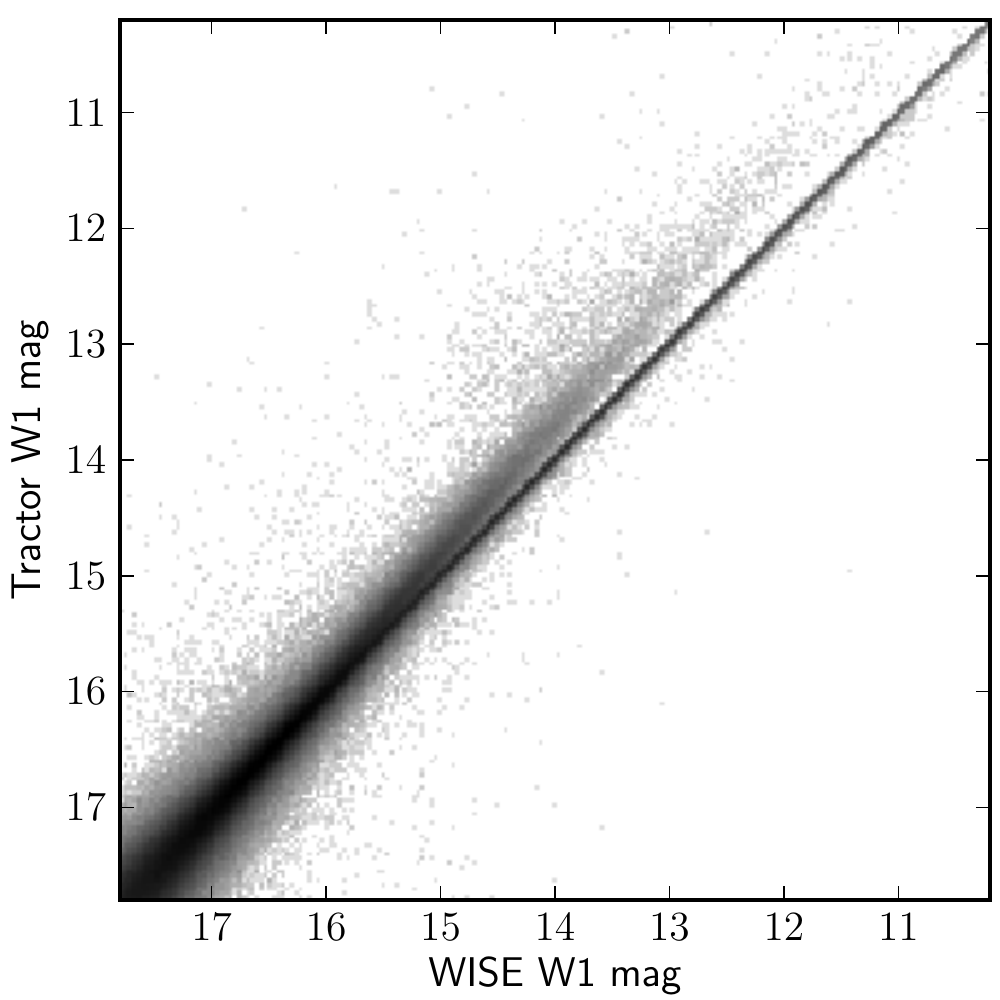} &
\includegraphics[width=0.33\textwidth]{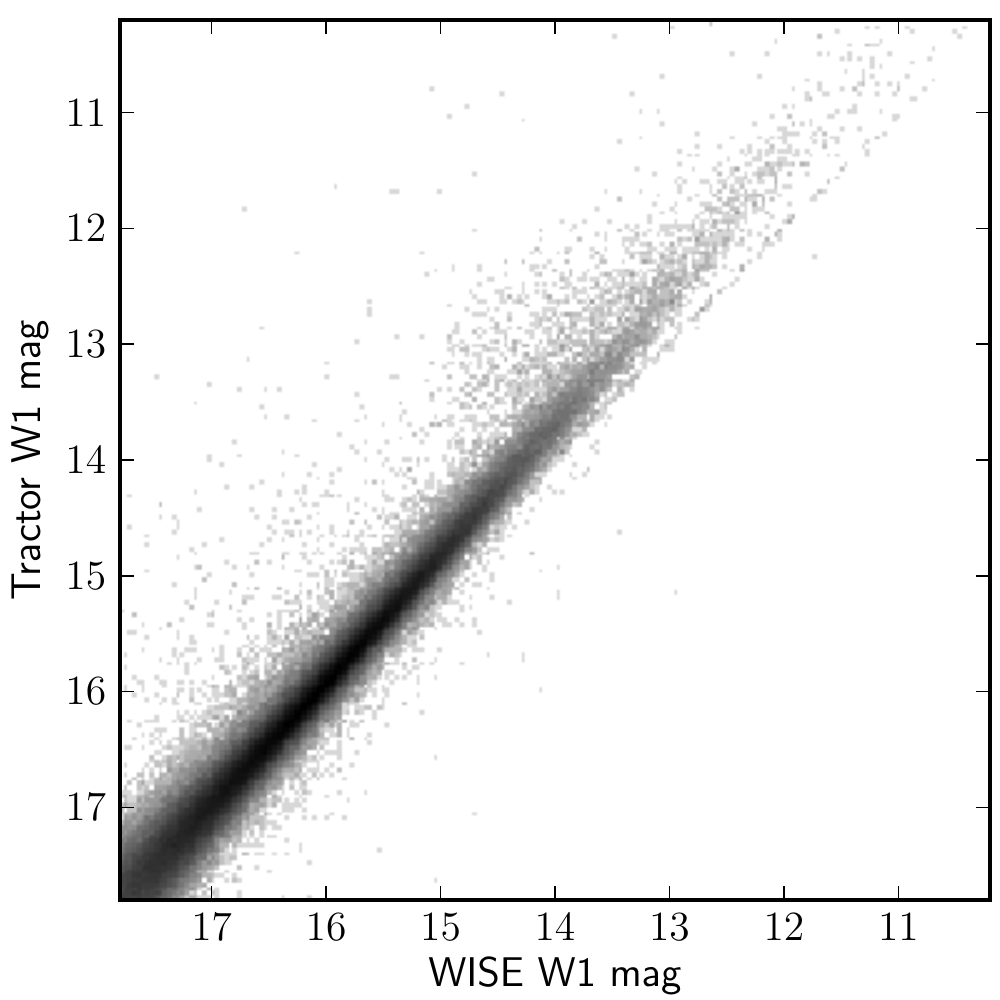} &
\includegraphics[width=0.33\textwidth]{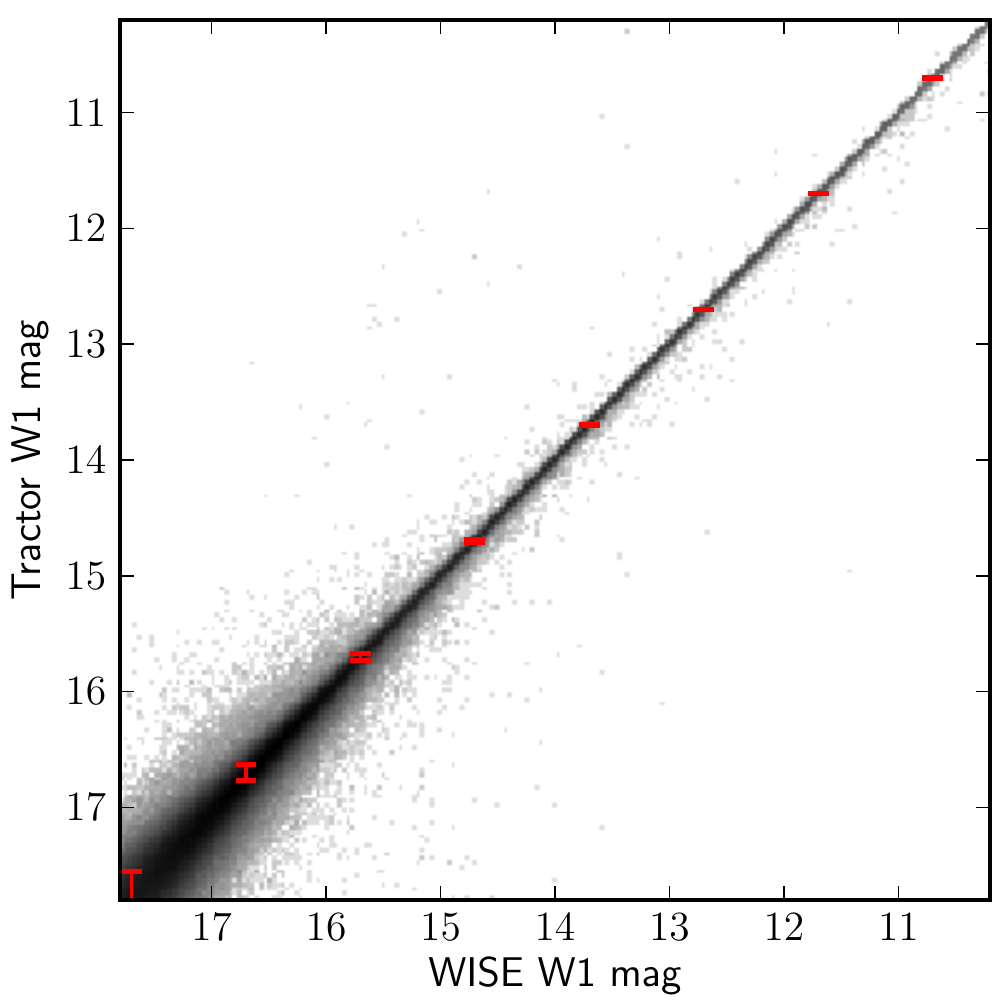}
\end{tabular}
\caption{Comparison of our forced photometry magnitudes and WISE
  AllWISE Release catalog magnitudes, in W1, for matched sources
  within 4 arcseconds, in $\sim 100$ square degrees of sky around
  RA,Dec = $(180, 45)$.  We show only unique (one-to-one) matches,
  since otherwise the \tractor\ photometry resolves sources that are
  blended in WISE.  \textbf{Left:} All sources; \textbf{Middle:} Sources
  identified in the SDSS imaging as galaxies, and not treated as point
  sources in our photometry; \textbf{Right:} sources identified by SDSS
  as point-like, plus nominally extended sources treated as point
  sources in our photometry.  The error bars shown are the median WISE
  catalog error bars per magnitude bin.  The WISE catalog magnitude
  entry we are plotting is \emph{w1mpro}, a point-source measurement;
  this explains why the \tractor\ measurements of galaxies tend to be
  brighter: we measure all the galaxy's flux, while the WISE catalog
  only measures the fraction in the point-like core.
  \label{fig:comp}}
\end{figure}

\begin{figure}
\begin{center}
\begin{tabular}{@{}c@{\hspace{3em}}c@{}}
W1 & W2 \\
\includegraphics[width=0.4\textwidth]{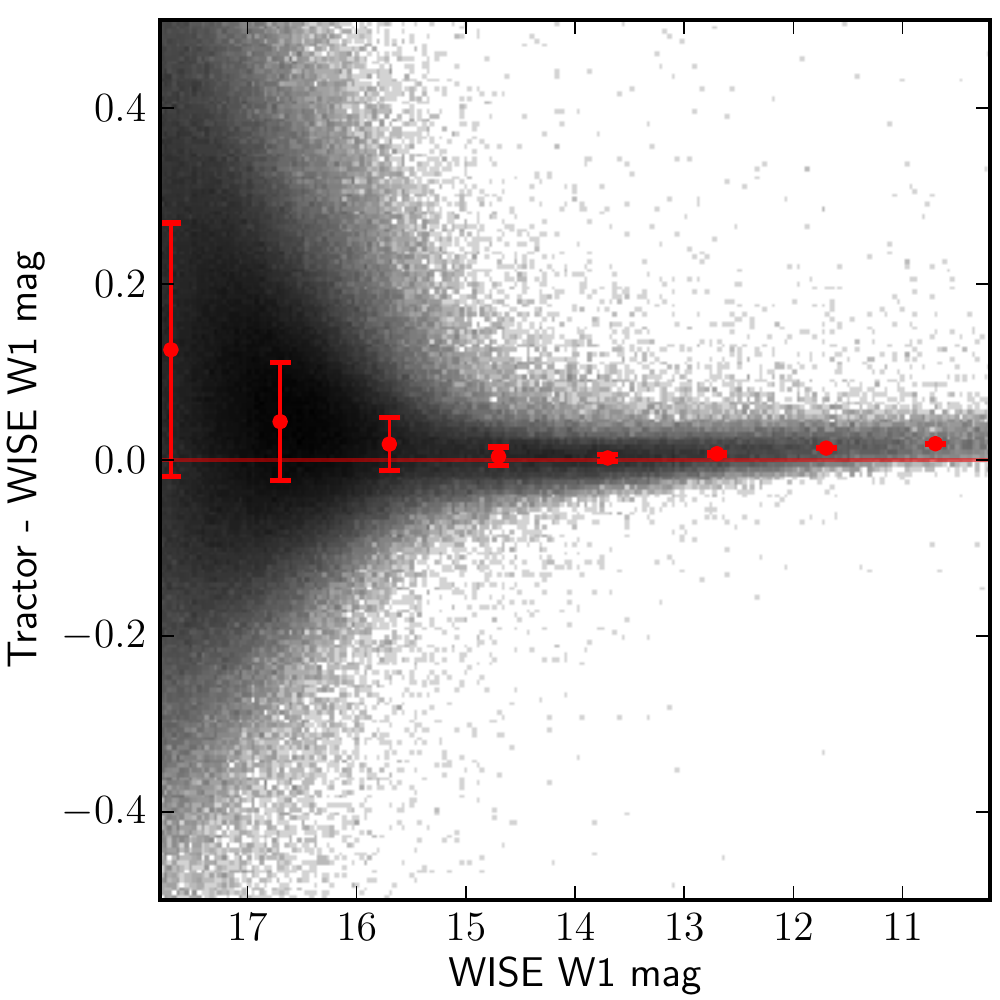} &
\includegraphics[width=0.4\textwidth]{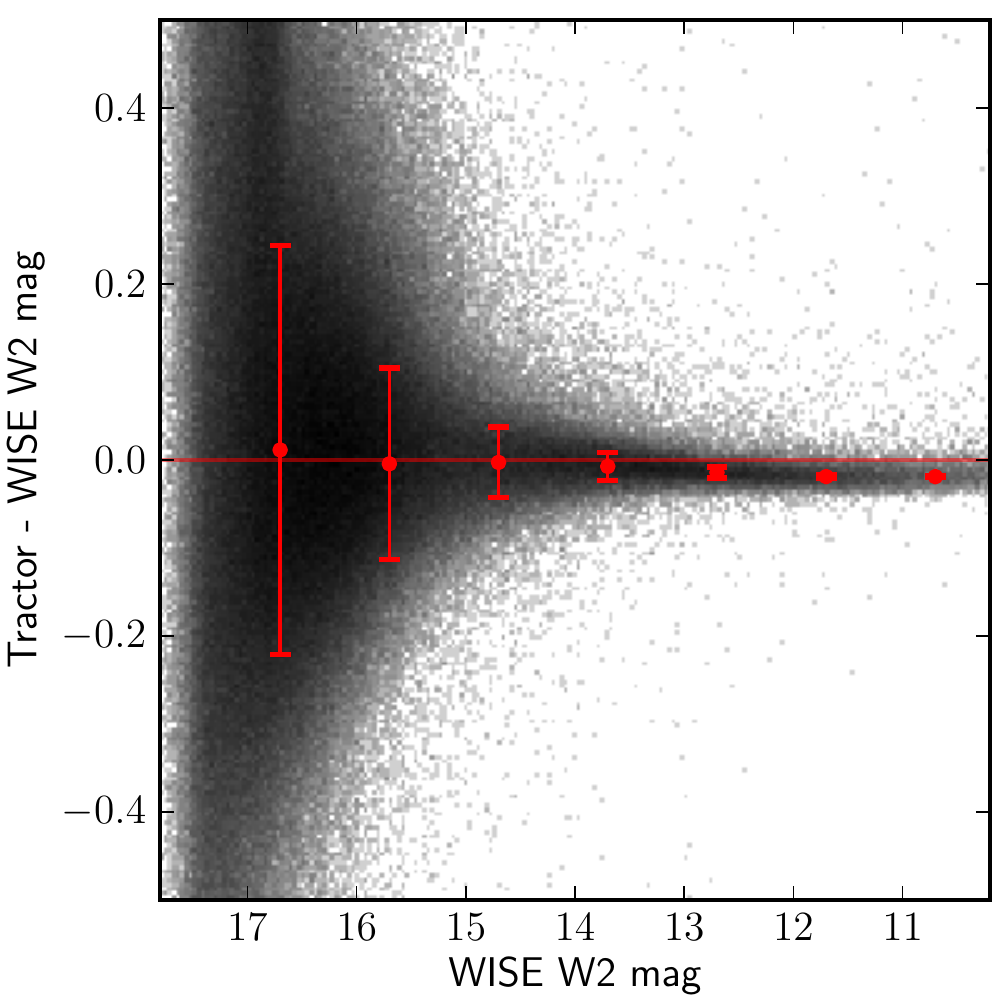} \\
W3 & W4 \\
\includegraphics[width=0.4\textwidth]{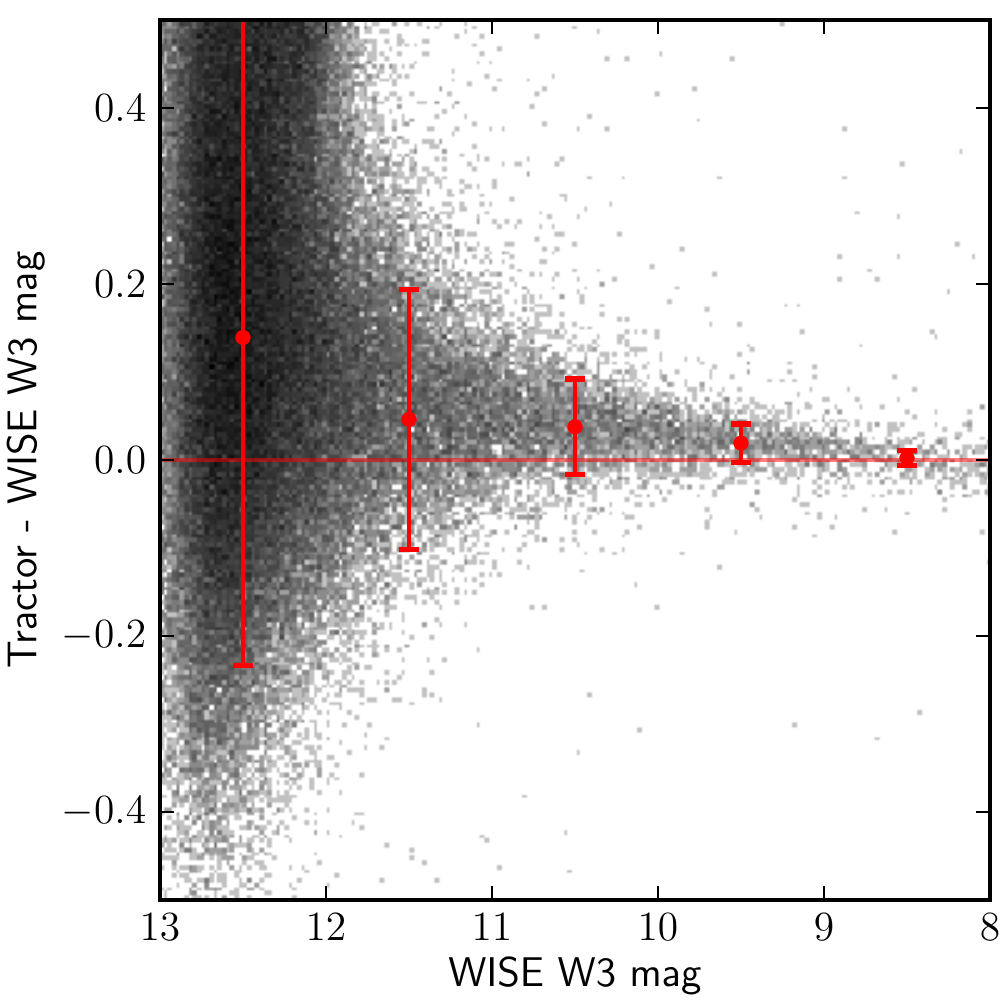} &
\includegraphics[width=0.4\textwidth]{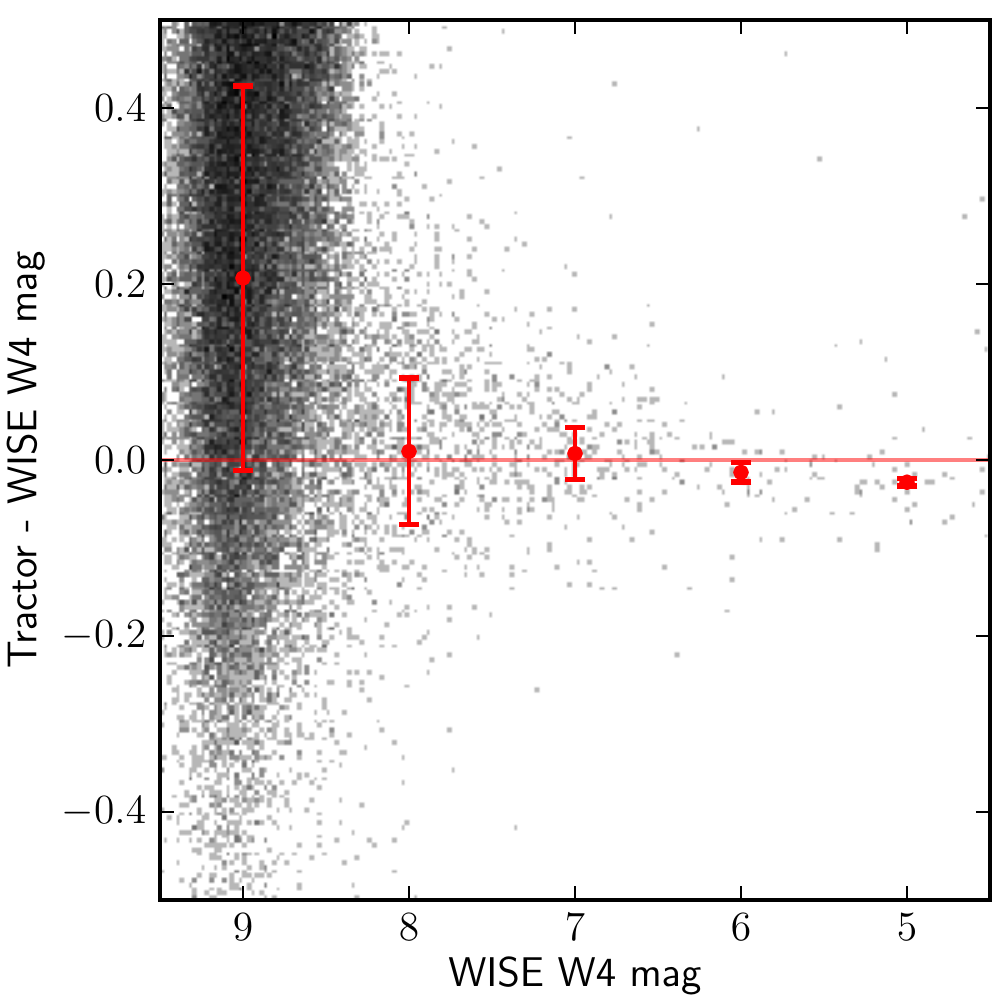}
\end{tabular}
\end{center}
\caption{Comparison of our forced photometry magnitudes and WISE
  AllWISE Release catalog magnitudes, for sources treated as point
  sources in our photometry.  These are sources in the same $\sim 100$
  square degrees in \figref{fig:comp}.  The slight tilts seen are
  comparable in magnitude to the differences between the All-Sky and
  AllWISE releases, and may be due to differences in photometric
  calibration between the releases.
  \label{fig:comp2}}
\end{figure}

\begin{figure}
\begin{center}
\begin{tabular}{@{}c@{}c@{}}
  \includegraphics[width=0.48\textwidth]{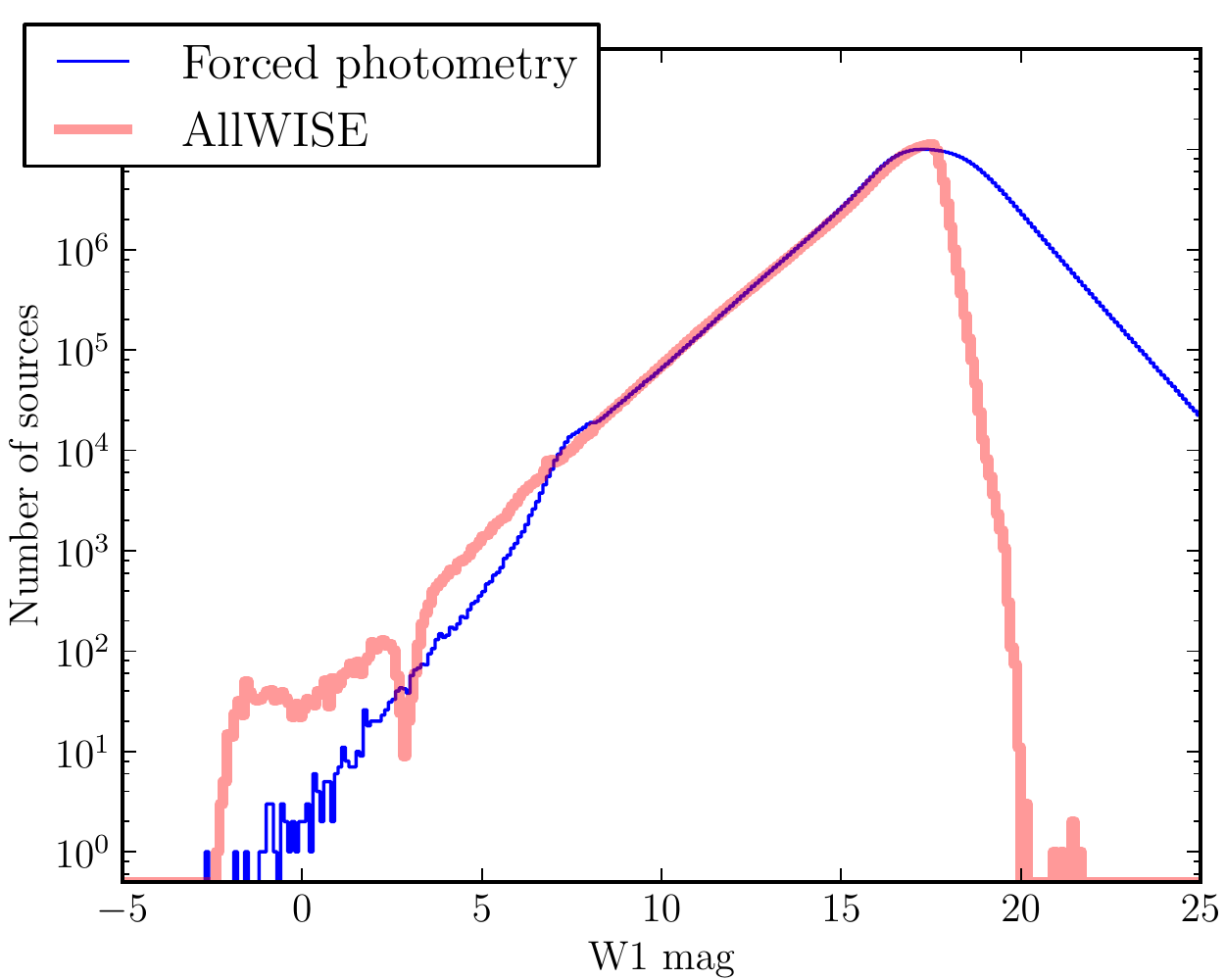} &
  \includegraphics[width=0.48\textwidth]{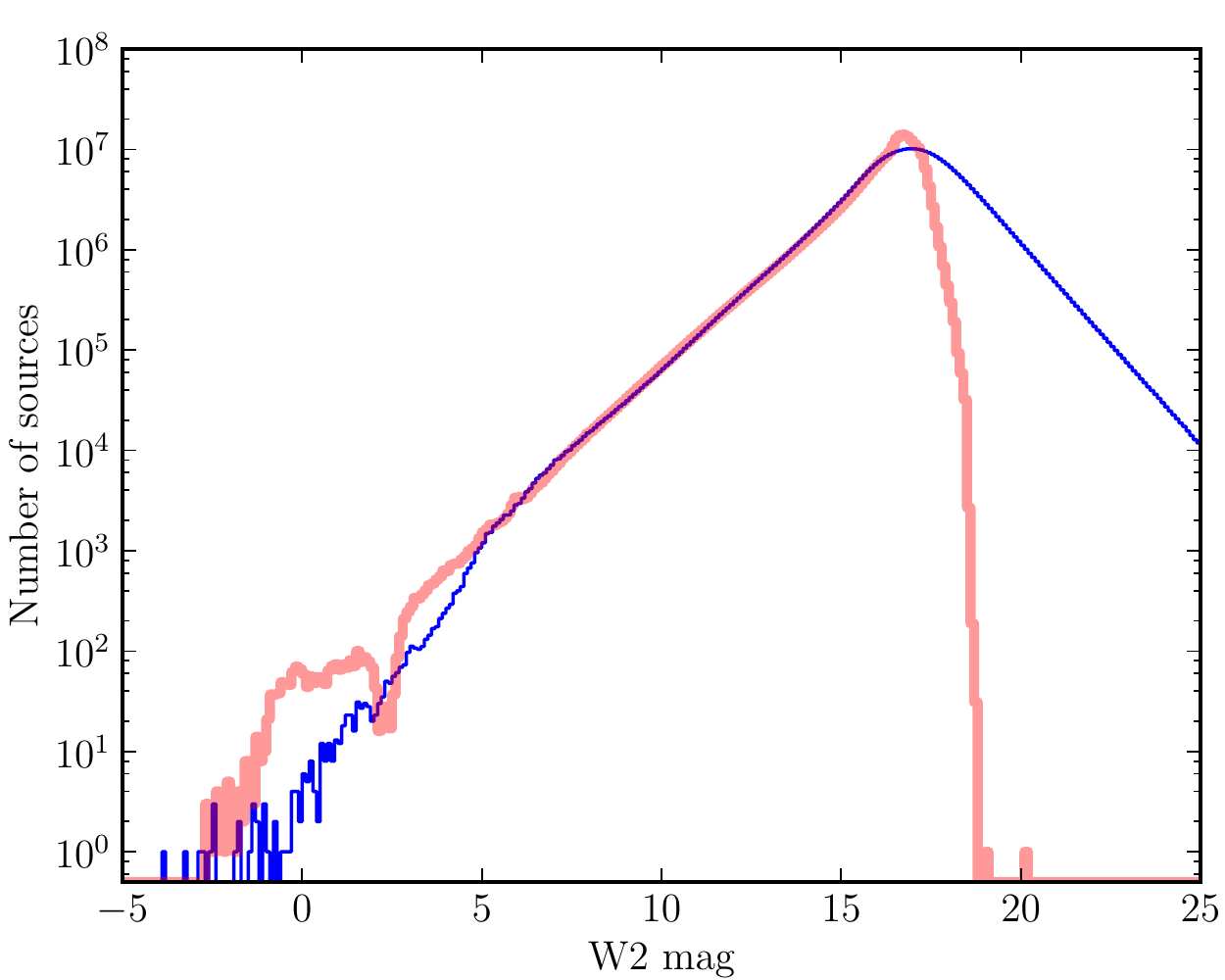} \\
  \includegraphics[width=0.48\textwidth]{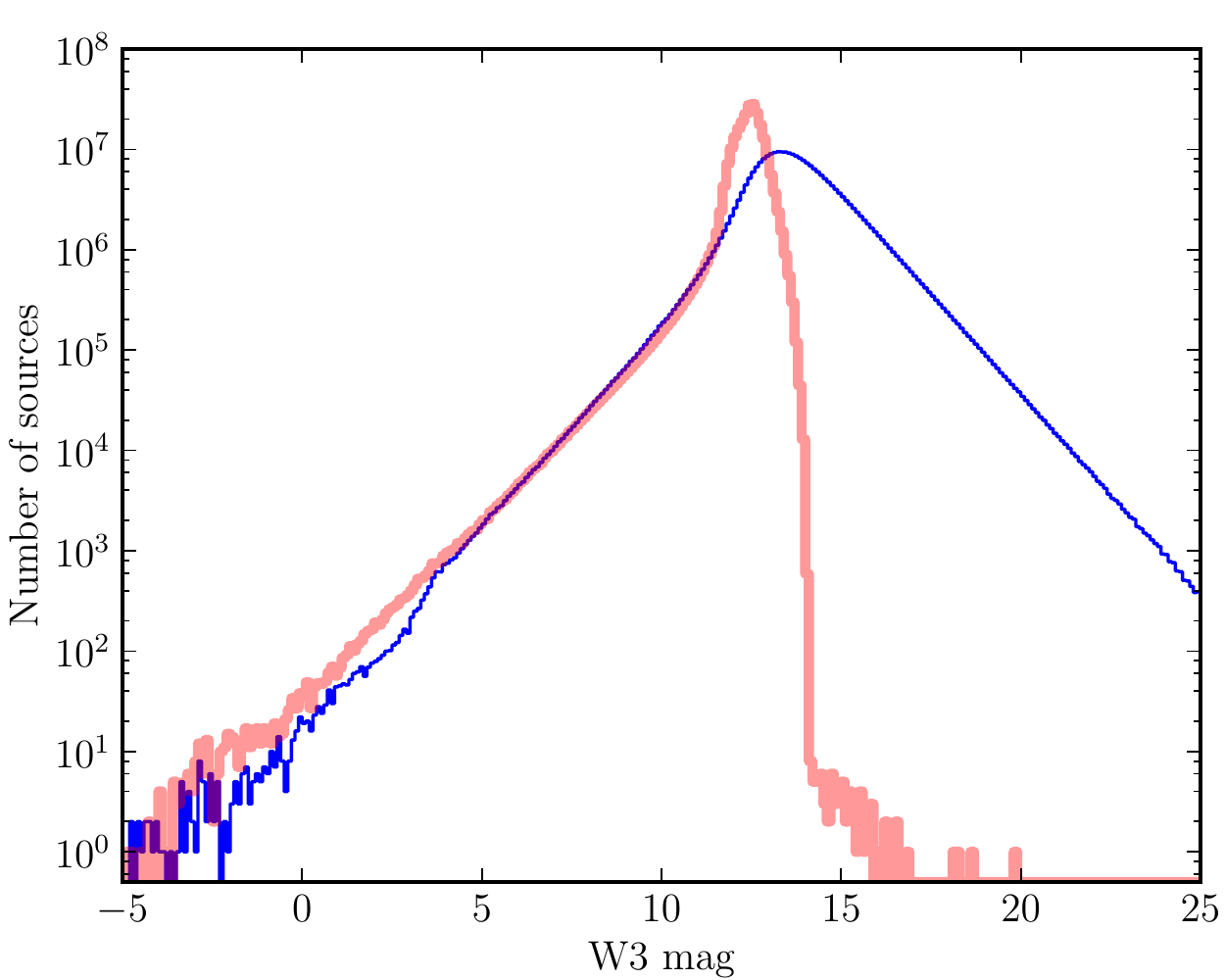} &
  \includegraphics[width=0.48\textwidth]{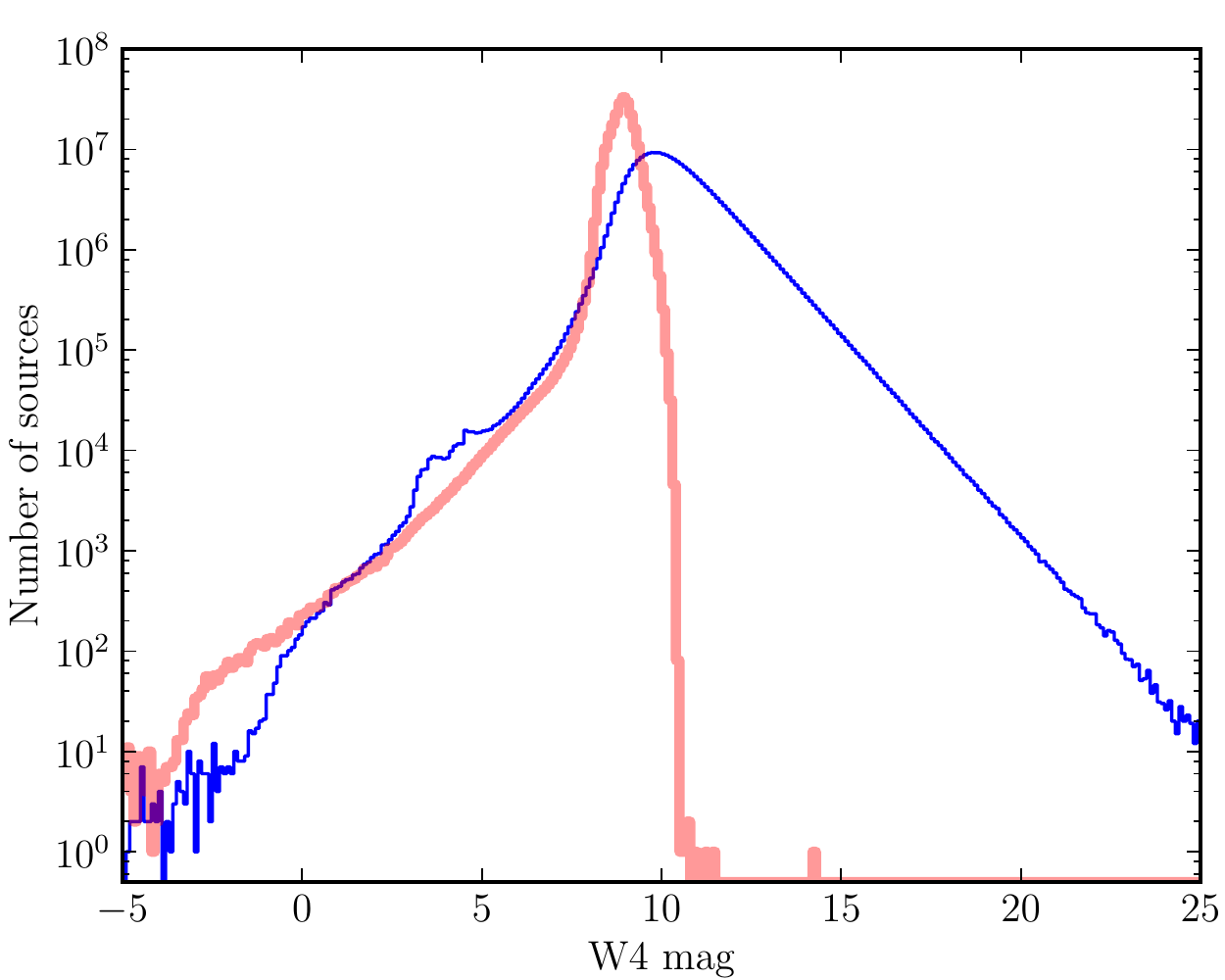}
\end{tabular}
\end{center}
\caption{Comparison of our forced photometry magnitudes and WISE
  AllWISE Release catalog magnitudes.  \textbf{Top-left}: W1.  Observe
  that above W1 mag $\sim$ 17, the AllWISE catalog detection
  efficiency drops sharply.  Our forced photometry results, in
  contrast, include faint measurements of many more sources (albeit at
  low signal-to-noise).  An unusual deficit of AllWISE sources around
  mag $\sim 3$ is apparent, while our forced photometry results show a
  rather smooth distribution.  The shape of the forced photometry
  distribution should be essentially the convolution of the SDSS optical
  detection efficiency and the optical--WISE colors.
  \textbf{Top-right}: W2 is similar.  \textbf{Bottom-left}: W3.  The
  AllWISE catalogs shows an upturn in number of sources of mag $11$ to
  $12$.  \textbf{Bottom-right}: W4 shows a similar effect.
  \label{fig:maghists}}
\end{figure}

\begin{figure}
\begin{center}
\begin{tabular}{@{}c@{}c@{}}
  \includegraphics[width=0.48\textwidth]{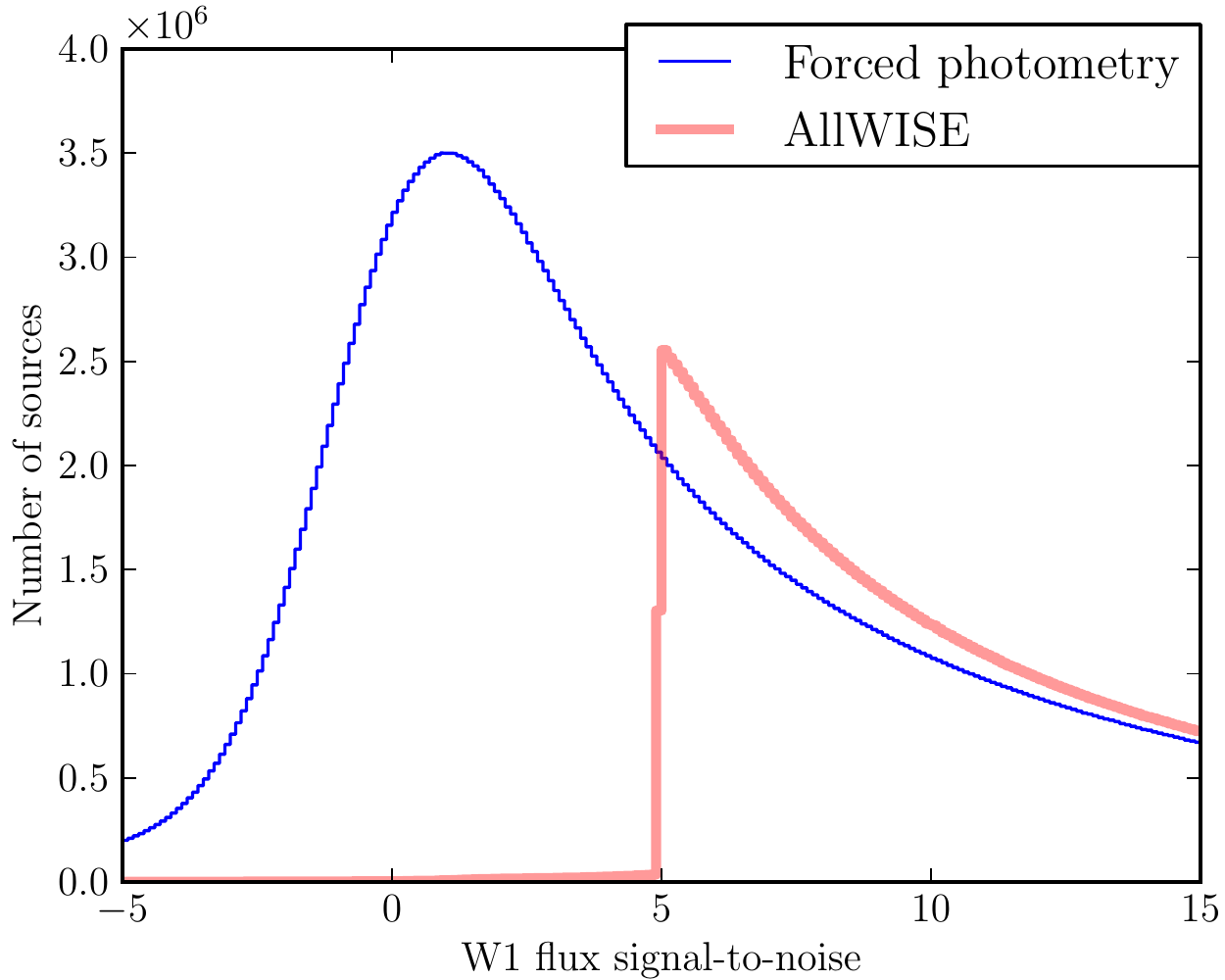} &
  \includegraphics[width=0.48\textwidth]{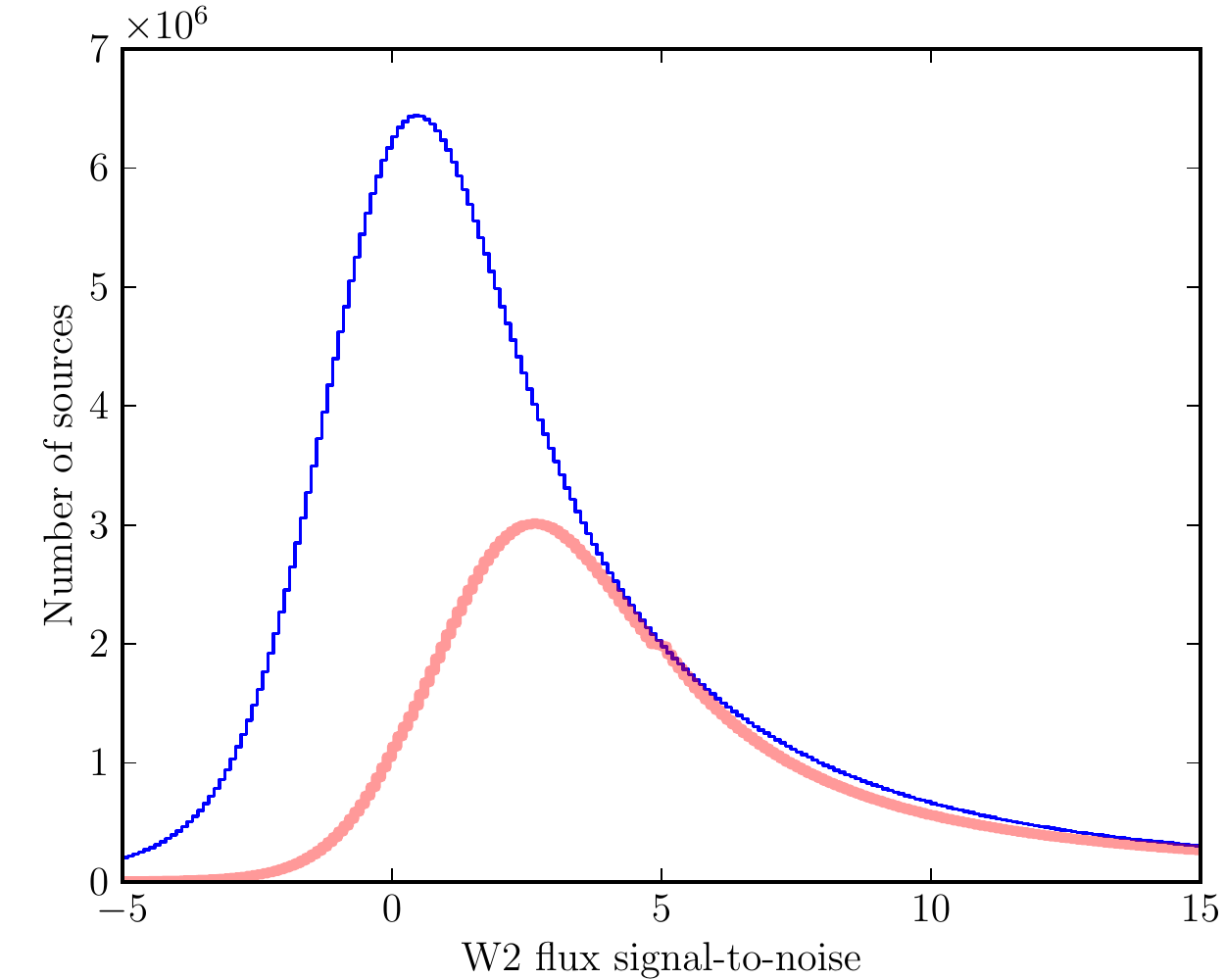} \\
  \includegraphics[width=0.48\textwidth]{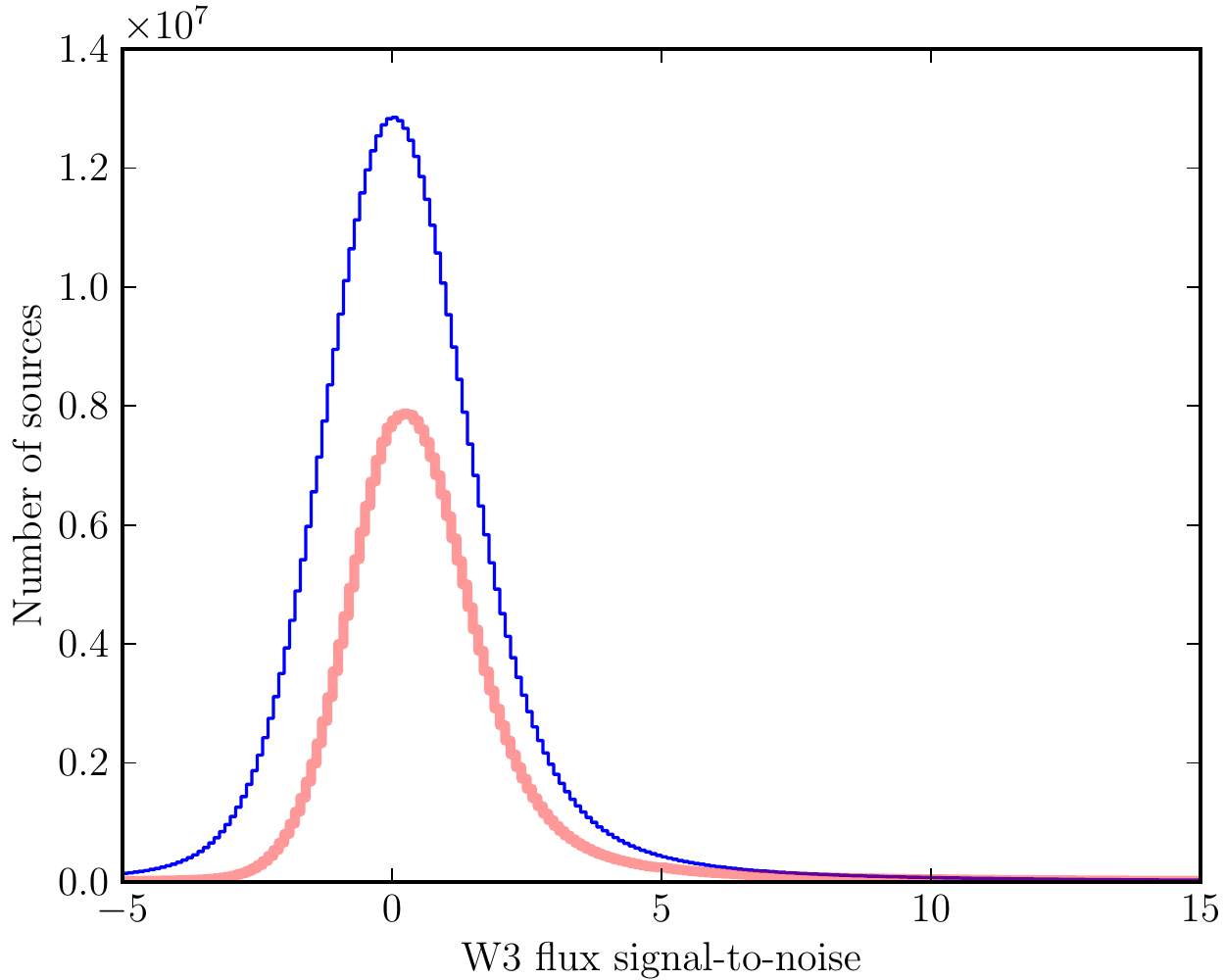} &
  \includegraphics[width=0.48\textwidth]{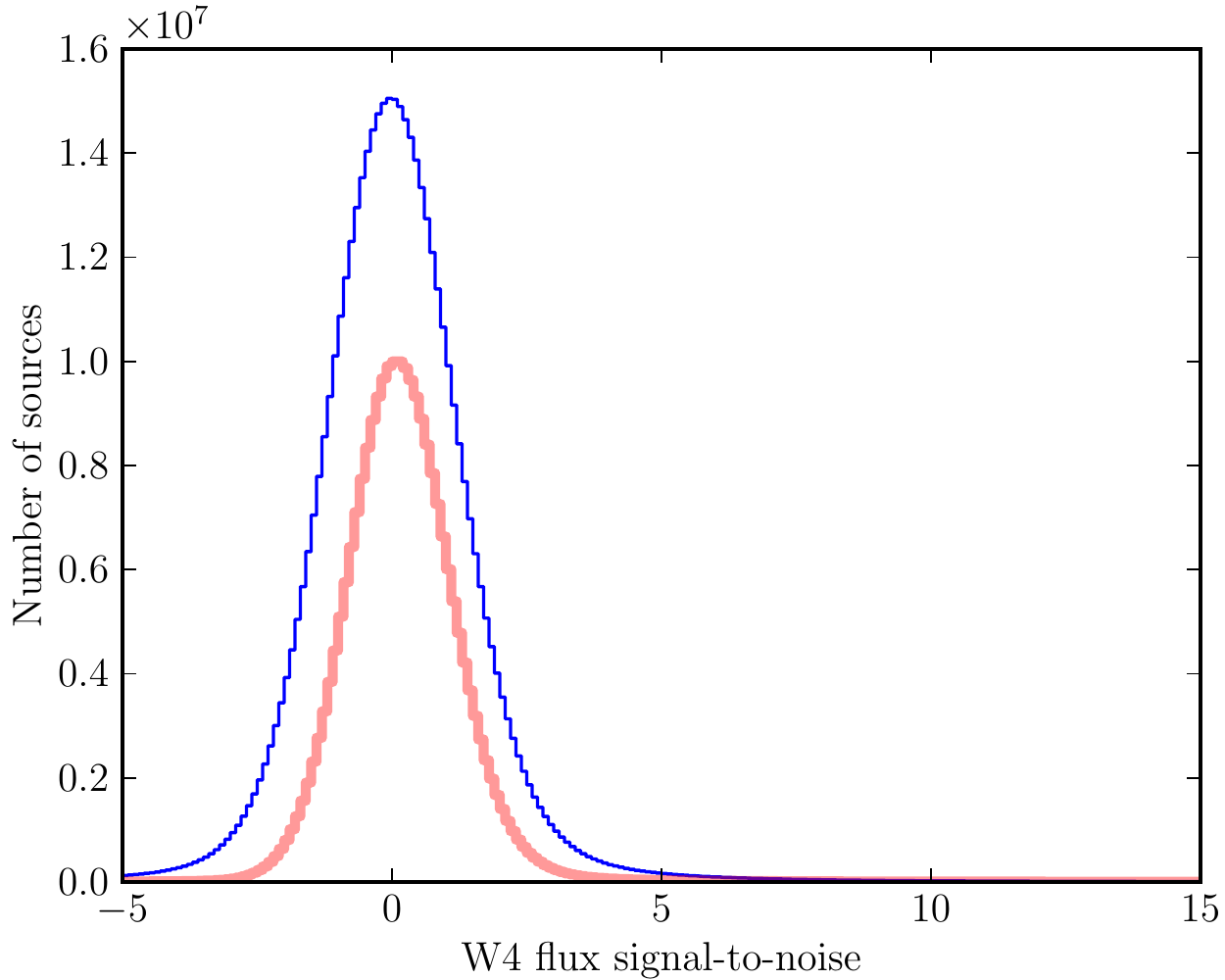}
\end{tabular}
\end{center}
\caption{Comparison of our forced photometry magnitudes and WISE
  AllWISE Release signal-to-noise distributions.  \textbf{Top-left}:
  W1.  Since WISE is most sensitive in W1, the AllWISE catalog is
  essentially W1-selected.  As a result, the W1 measurements rise
  sharply at the W1 detection threshold.  Our forced photometry
  results, in contrast, include measurements of sources below the
  detection threshold.  \textbf{Top-right}: W2.  Our results for
  bright sources follow those of the AllWISE catalog, but we include
  more faint measurements.  \textbf{Bottom-left}: W3 and
  \textbf{Bottom-right}: W4 are similar.
  \label{fig:snhists}}
\end{figure}

\begin{figure}
\begin{center}
  \begin{tabular}{@{}c@{}}
    AllWISE \\
    \includegraphics[width=\textwidth]{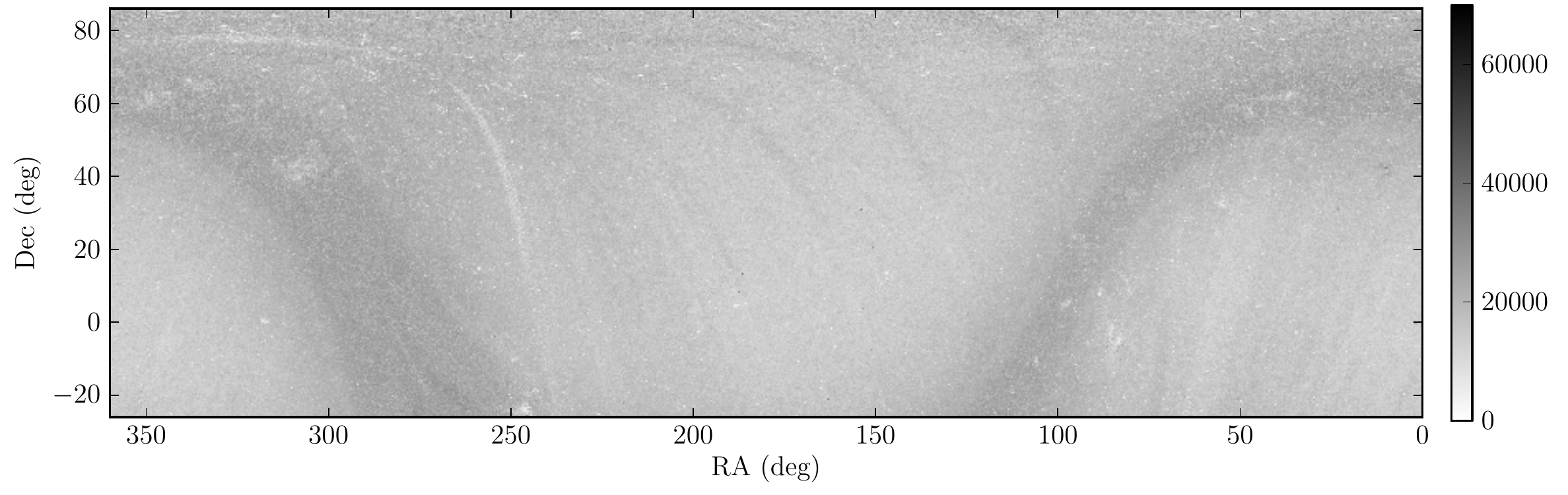} \\
    Treated as point sources \\
    \includegraphics[width=\textwidth]{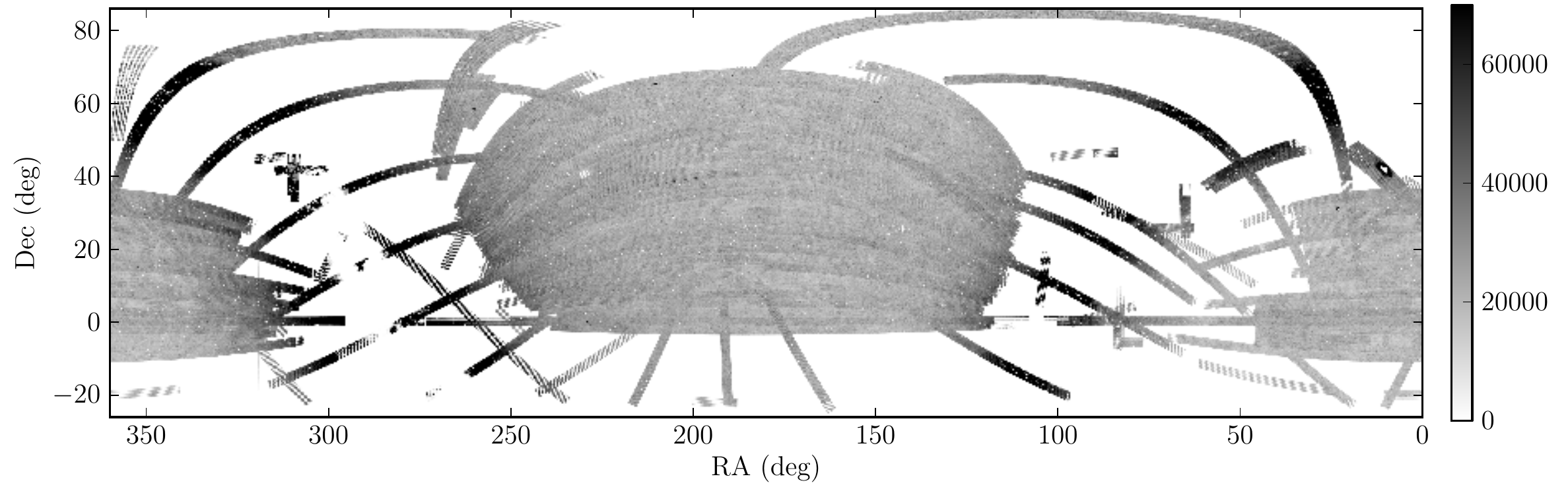} \\
    Treated as galaxies \\
    \includegraphics[width=\textwidth]{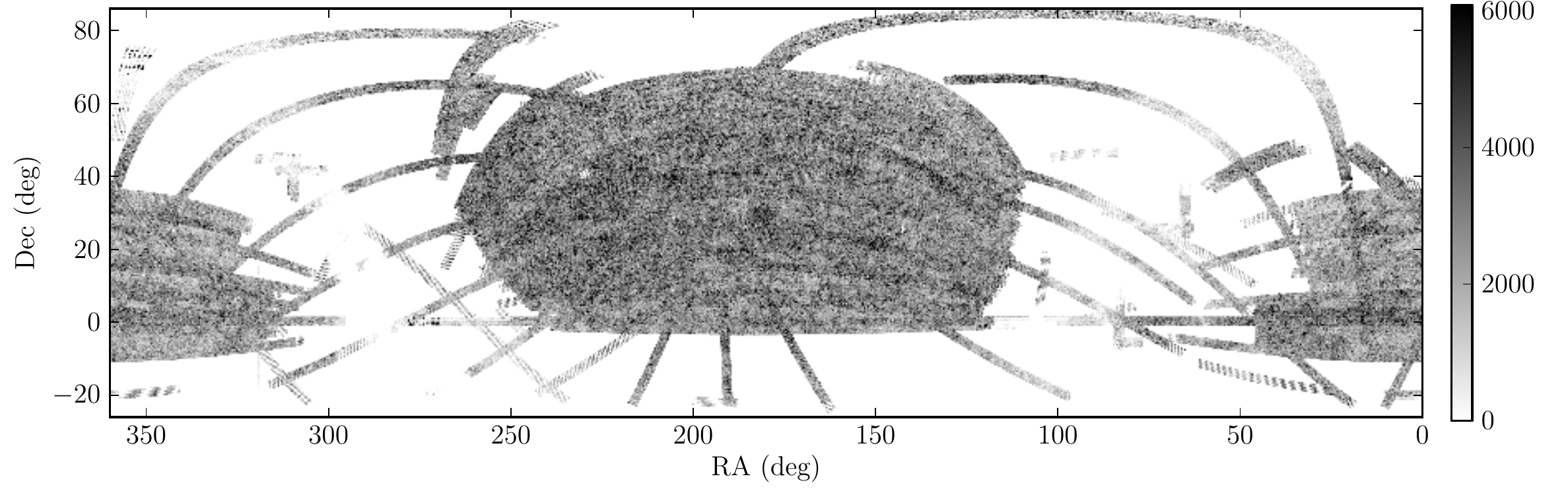}
  \end{tabular}
\end{center}
\caption{Spatial distribution of measurements.  \textbf{Top}: objects
  in the AllWISE catalog.  \textbf{Center}: objects in our forced
  photometry results that were treated as point sources.
  \textbf{Bottom}: sources we treated as galaxies.  The units are
  sources per square degree; note that we only treat $\sim 10\%$ of
  the sources as galaxies.
  \label{fig:maps}}
\end{figure}

Comparisons between our results and the ``official'' WISE catalog are
shown in \figurename s \ref{fig:comp}, \ref{fig:comp2},
\ref{fig:maghists}, \ref{fig:snhists}, and \ref{fig:maps}.  For isolated point
sources, our results are consistent to within about 0.03 mag.  A
slight tilt is evident, similar to the tilt seen in the comparison of
the All-Sky and AllWISE releases.  We expect this is due to
photometric calibration differences between the All-Sky and AllWISE
releases.  Unfortunately, the AllWISE ``level 1b'' calibrated
exposures have not been released, so improving this effect is beyond
the scope of this work.

\subsection{Sources undetected in the WISE catalog}

\begin{figure}
\begin{center}
  \begin{tabular}{@{}rcc@{}}
    & \hspace{1em} \makebox[0pt][c]{Matching} \hspace{0.21\textwidth}
    \makebox[0pt][c]{Forced Photometry} & \\
    \raisebox{0.12\textheight}{\footnotesize{$13 < \textrm{W1} < 17$}} &
    \includegraphics[height=0.23\textheight]{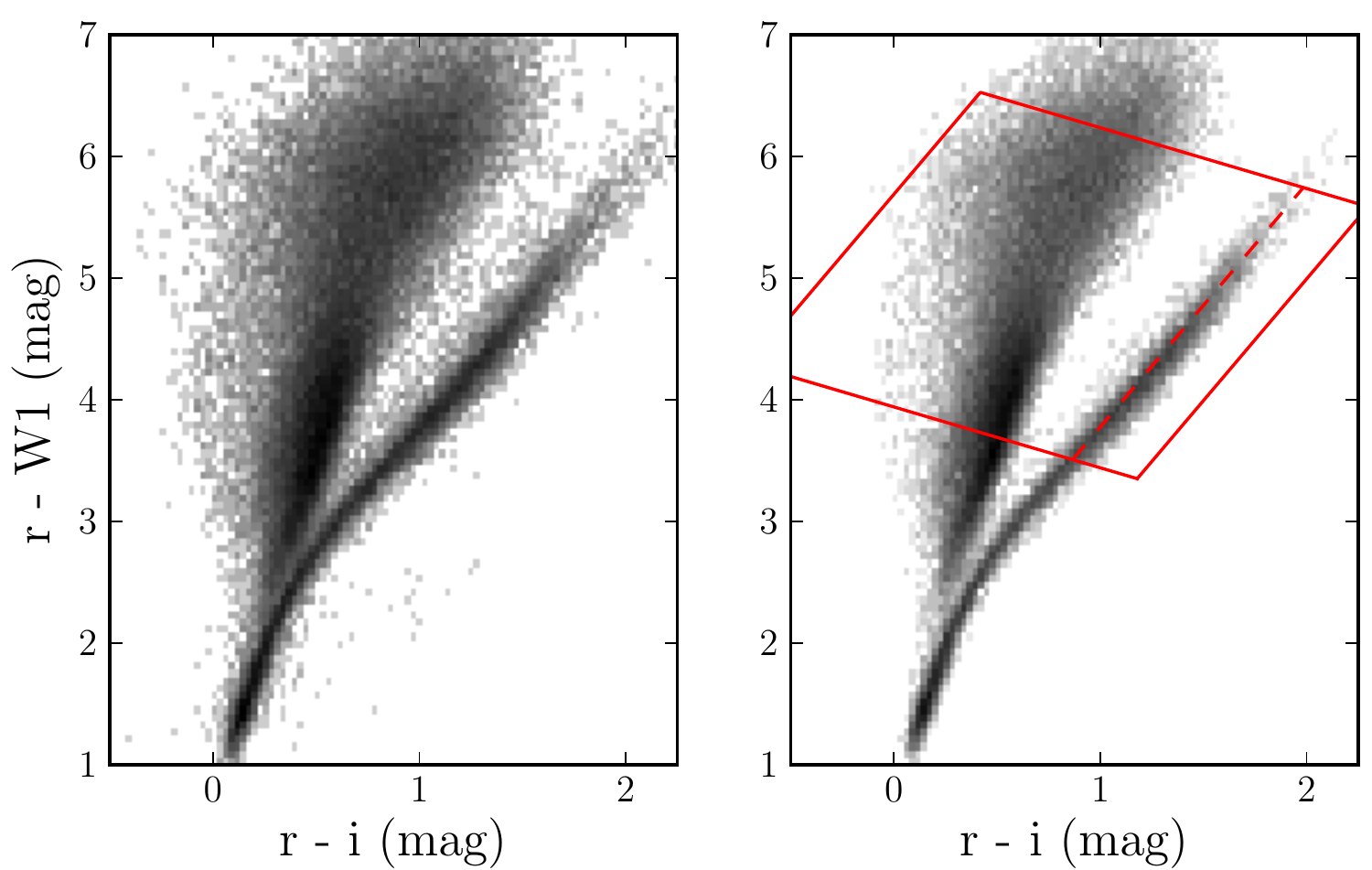} &
    \includegraphics[height=0.23\textheight]{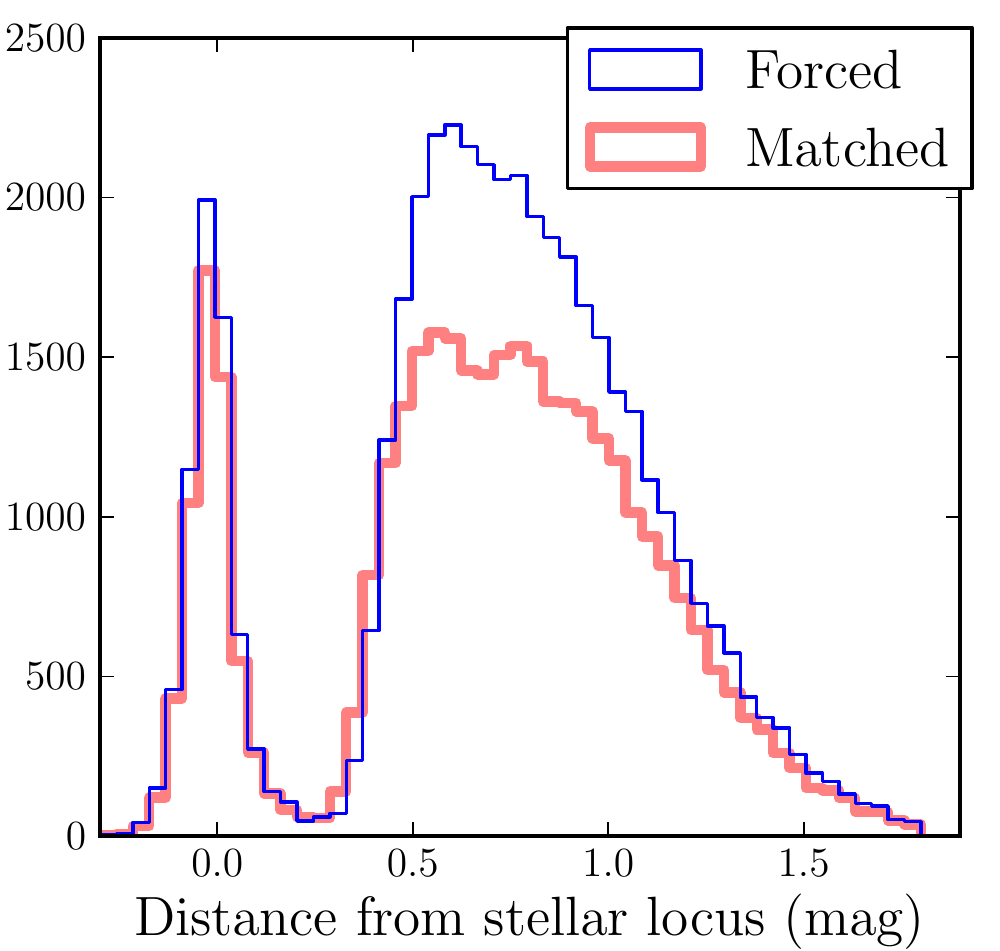} \\
    \raisebox{0.12\textheight}{\footnotesize{$17 < \textrm{W1} < 18$}} &
    \includegraphics[height=0.23\textheight]{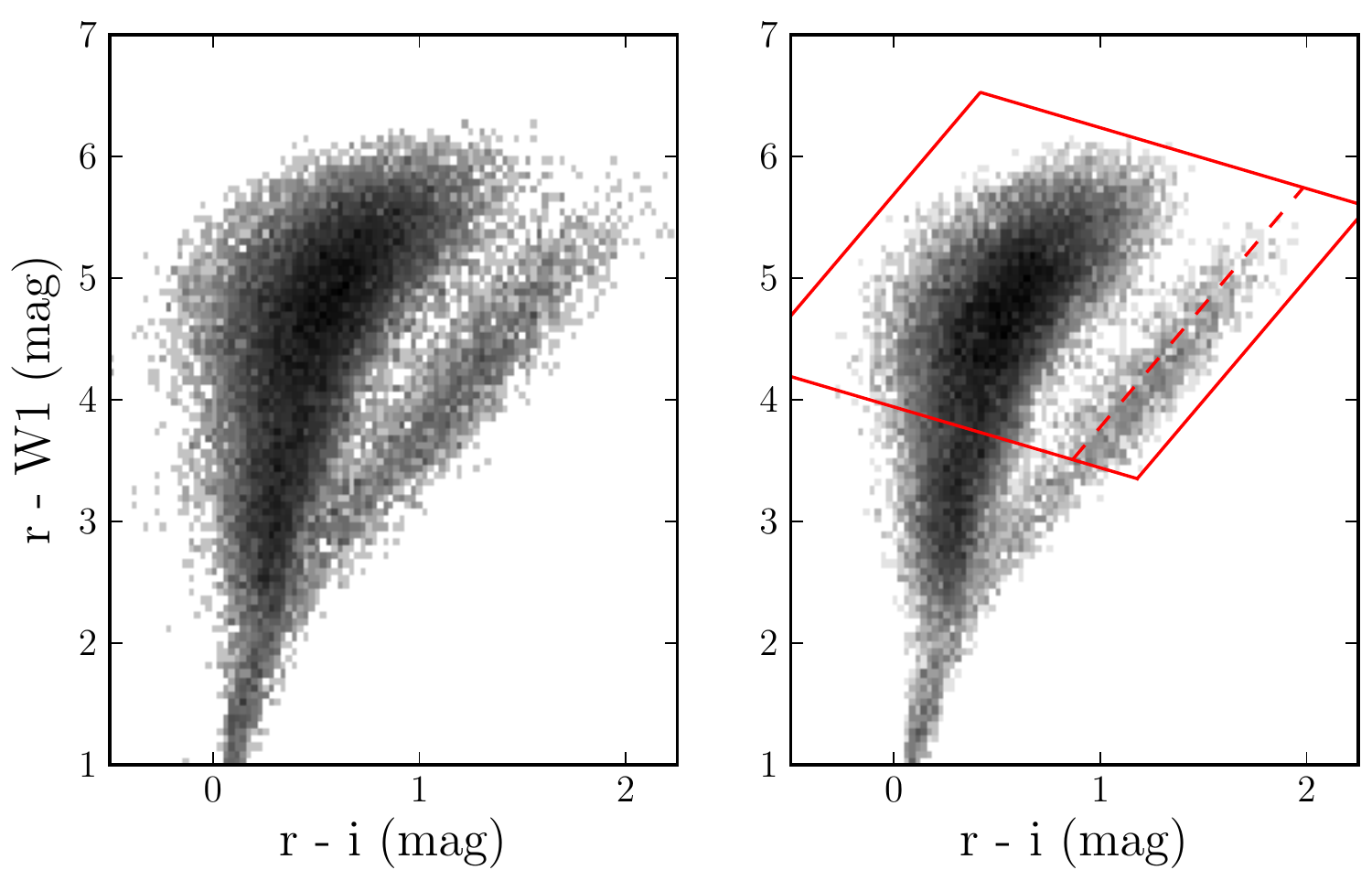} &
    \includegraphics[height=0.23\textheight]{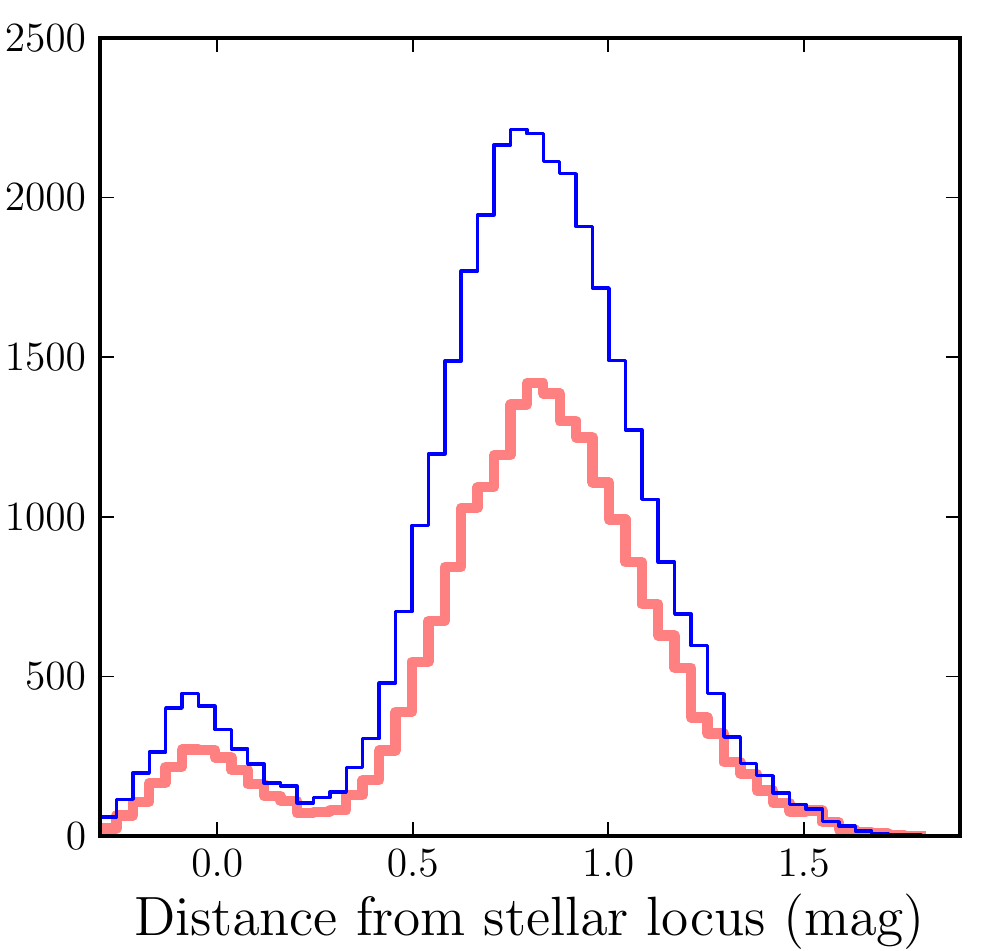} \\
    \raisebox{0.12\textheight}{\footnotesize{$18 < \textrm{W1} < 18.5$}} &
    \includegraphics[height=0.23\textheight]{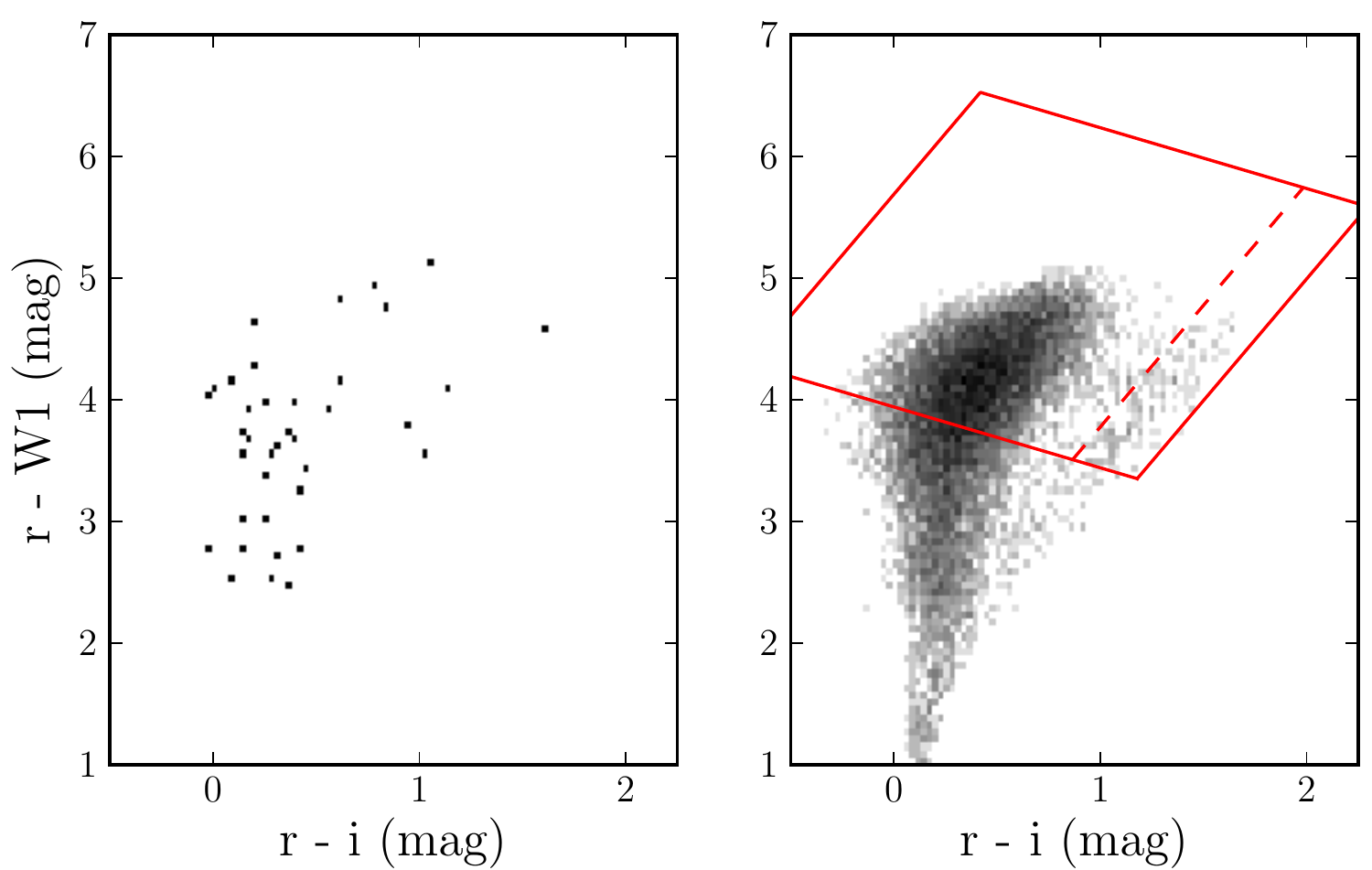} &
    \includegraphics[height=0.23\textheight]{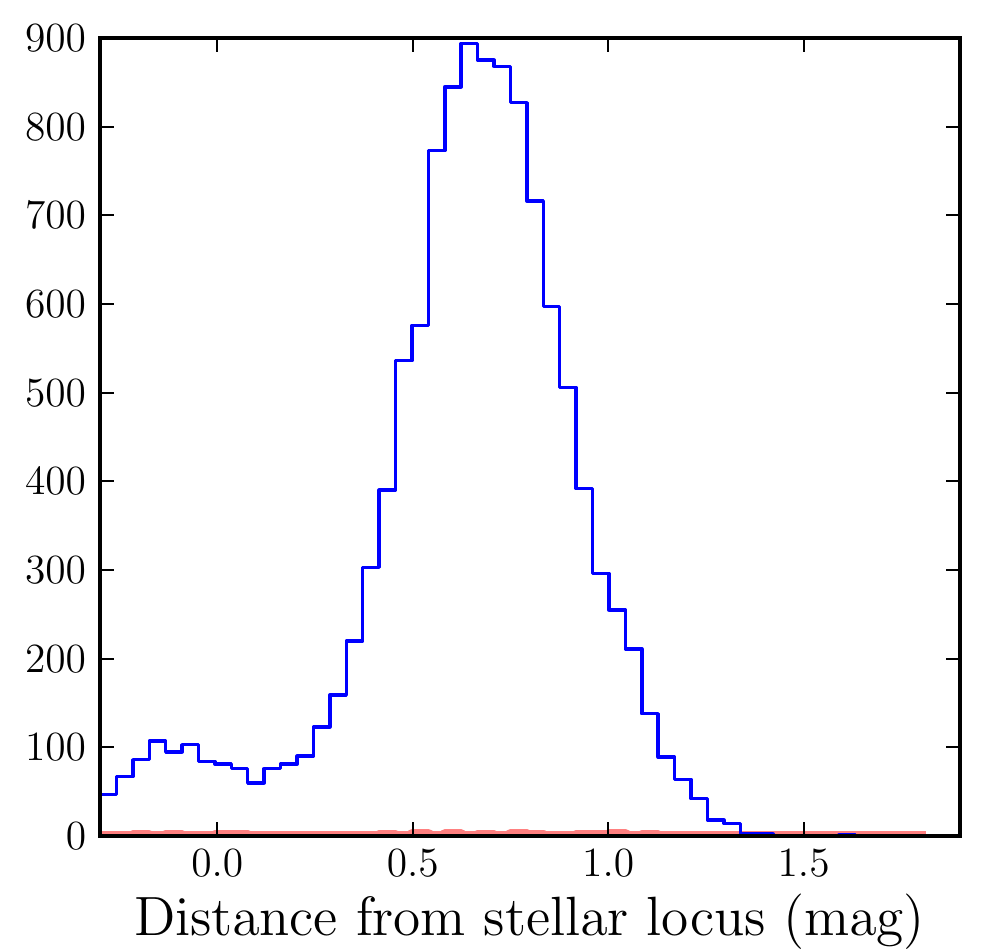}
  \end{tabular}
\end{center}
\caption{Comparison between our forced-photometry measurements and
  traditional astrometric matching to the WISE catalog.  The $r - W1$
  vs $r - i$ color-color plane is used for the selection of luminous
  red galaxy (LRG) targets in the SDSS-III/SEQUELS survey.
  \textbf{Top}: bright sources.  \textbf{Middle}: moderate sources.
  \textbf{Bottom}: faint sources in W1.  The scatterplots show the
  density of sources in color-color space, using astrometric matching
  (left) or our forced photometry results (middle).  The histogram
  (right) shows the distance of sources from the approximate stellar
  locus location (dashed line), within the indicated box.  For the
  bright sources, the results are similar, but we measure overall more
  sources thanks to the resolving power of SDSS.  At moderate
  brightness, we again get similar results.  The bump of sources near
  the stellar locus is broader due to photometric errors.  At faint
  brightness levels, the sources are not detected in WISE alone, so do
  not appear in the WISE catalog and cannot be matched.  In contrast,
  our photometric measurements remain reasonable below the WISE
  detection threshold and allow reliable separation of stars and LRG
  targets.
\label{fig:lrg}}
\end{figure}

\Figref{fig:lrg} shows a comparison between our forced-photometry
results as compared to a traditional approach of astrometric matching
between the SDSS and WISE catalogs.  The traditional approach demands
that all sources be detected in both catalogs, while our approach
allows few-sigma WISE sources to be measured.  For some science cases,
these few-sigma measurements can be very useful, either individually
or in stacking analyses.  In addition, we photometer more source
overall, since often SDSS resolves nearby sources that are blended in
WISE.

\subsection{Artifacts}

\begin{figure}
  \newlength{\figw}
  \setlength{\figw}{0.27\textwidth}
  \newcommand{\spc}{\hspace*{-0.03\textwidth}}
  \begin{center}
    \begin{tabular}{@{}c@{\spc}c@{\spc}c@{\spc}c@{}}
      WISE image & AllWISE Catalog & SDSS image & Forced Photometry \\
      %
      \includegraphics[width=\figw]{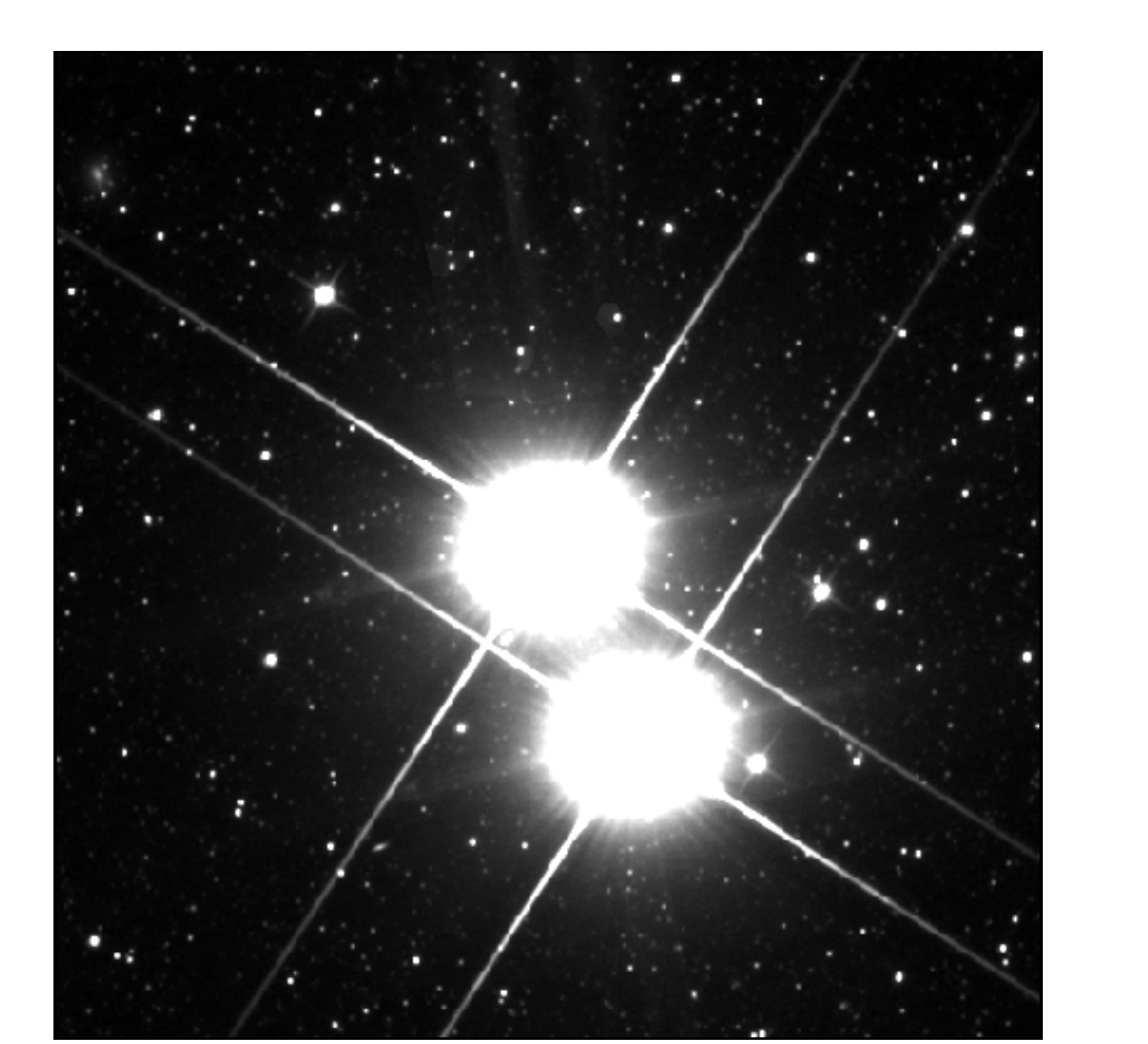} &
      \includegraphics[width=\figw]{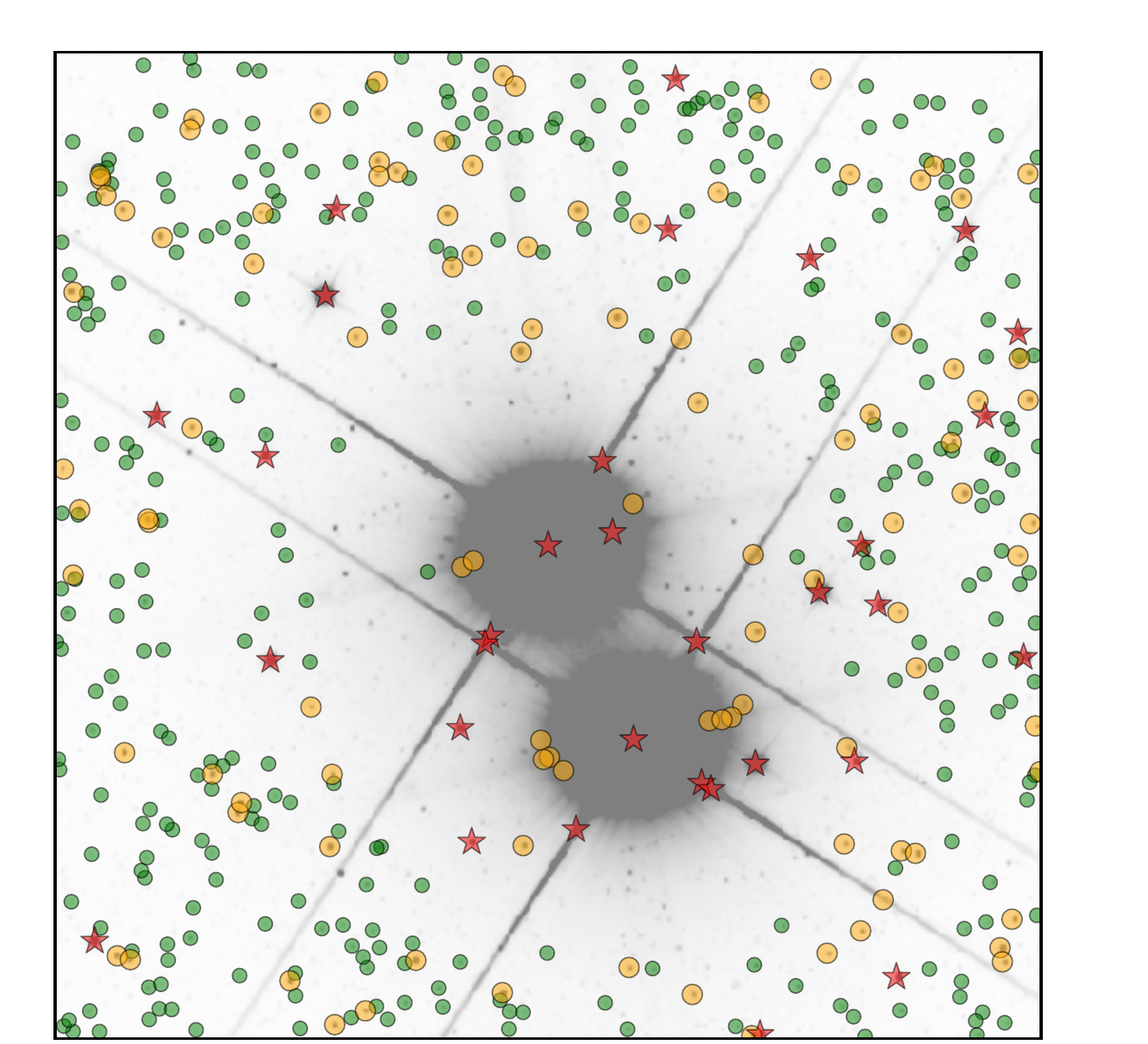} & 
      \includegraphics[width=\figw]{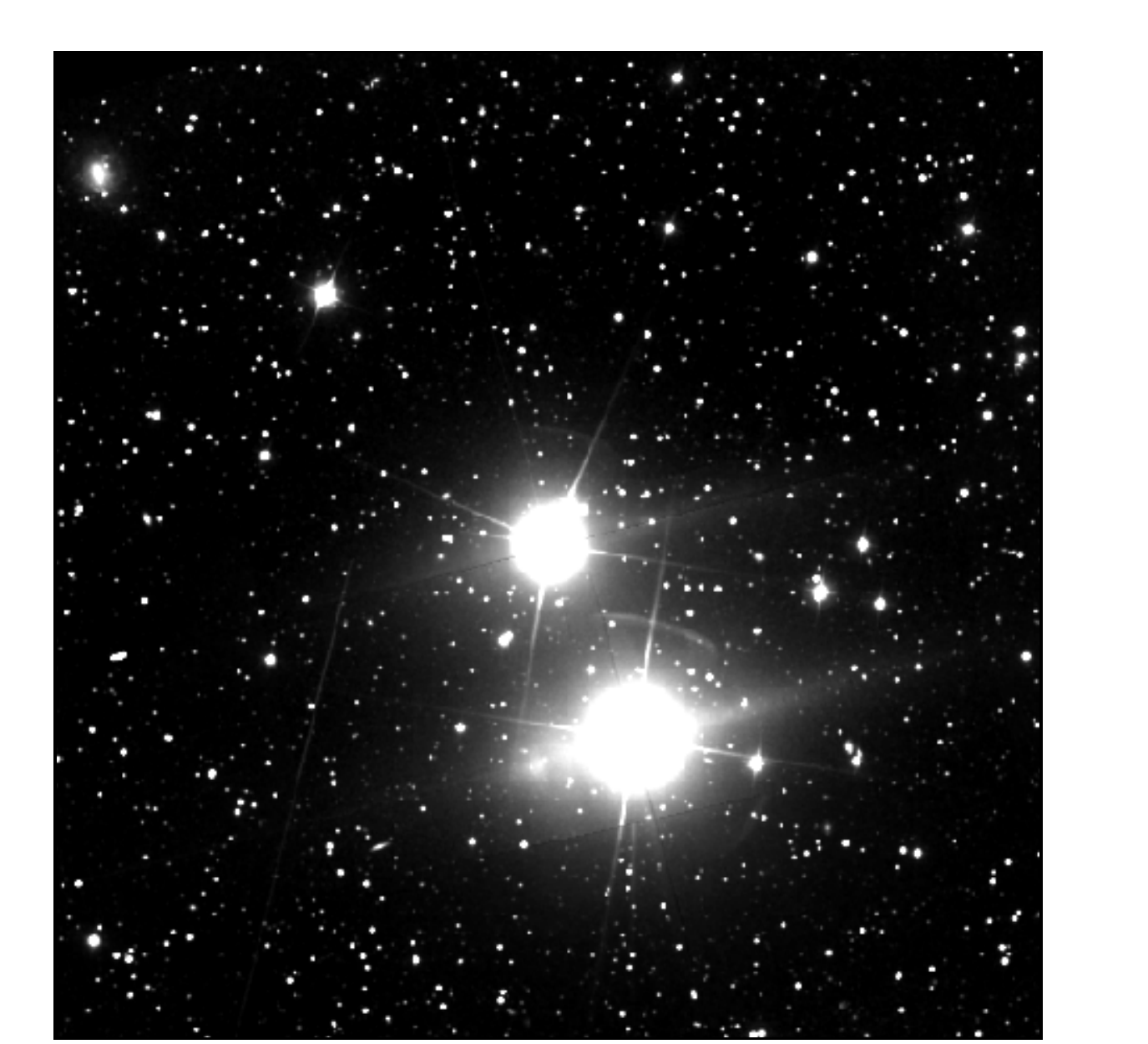} &
      \includegraphics[width=\figw]{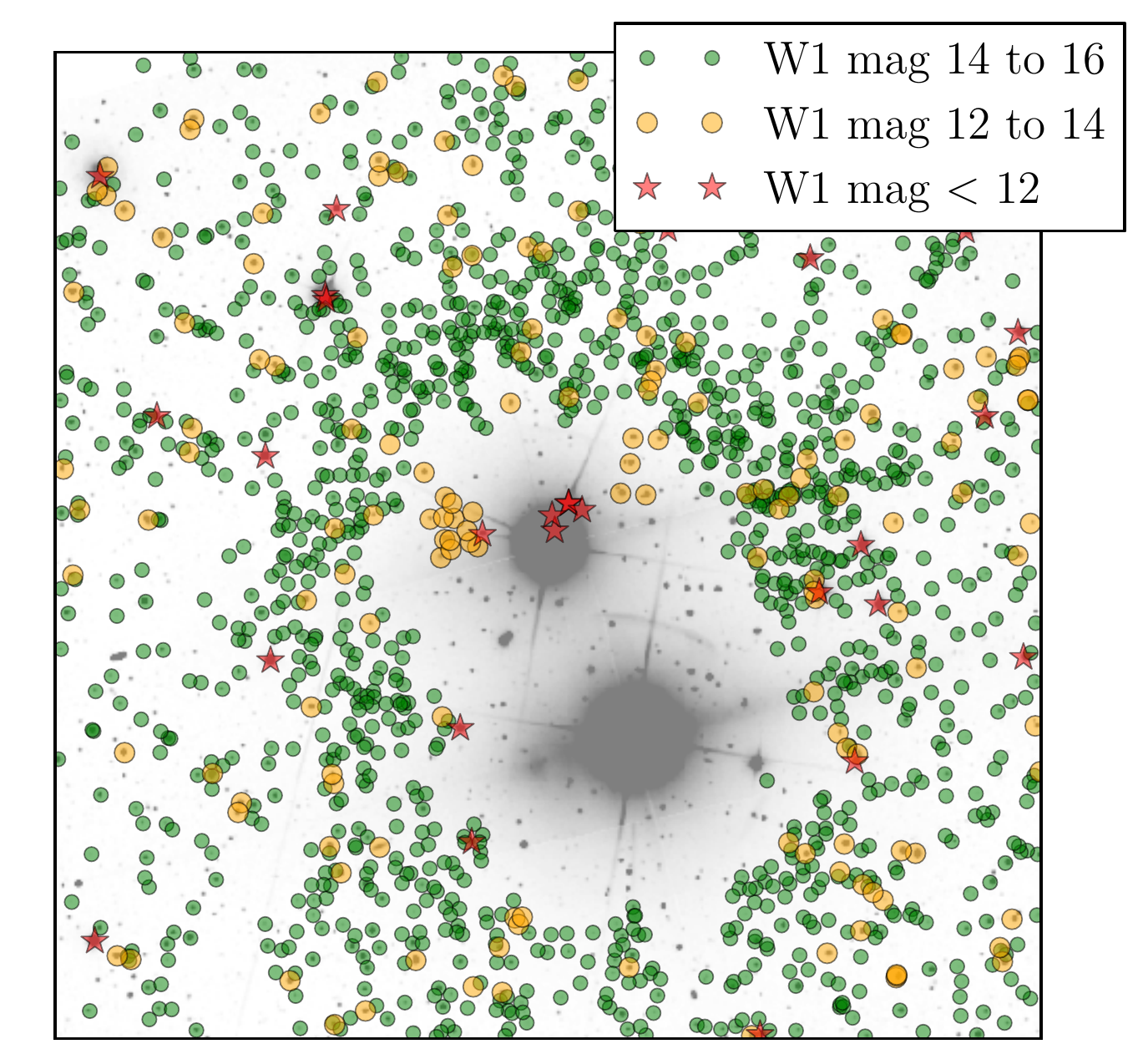} \\
      %
      \includegraphics[width=\figw]{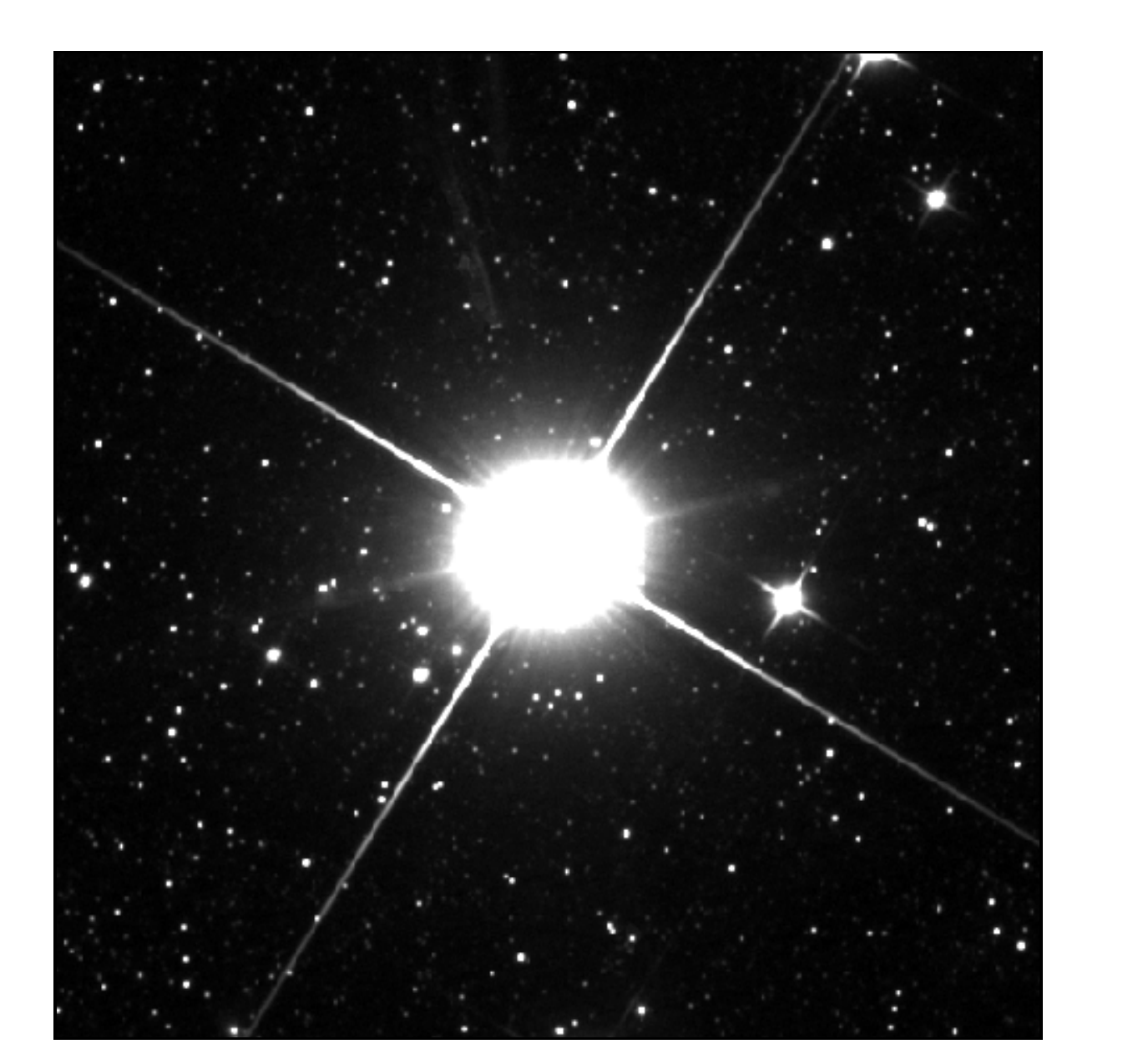} &
      \includegraphics[width=\figw]{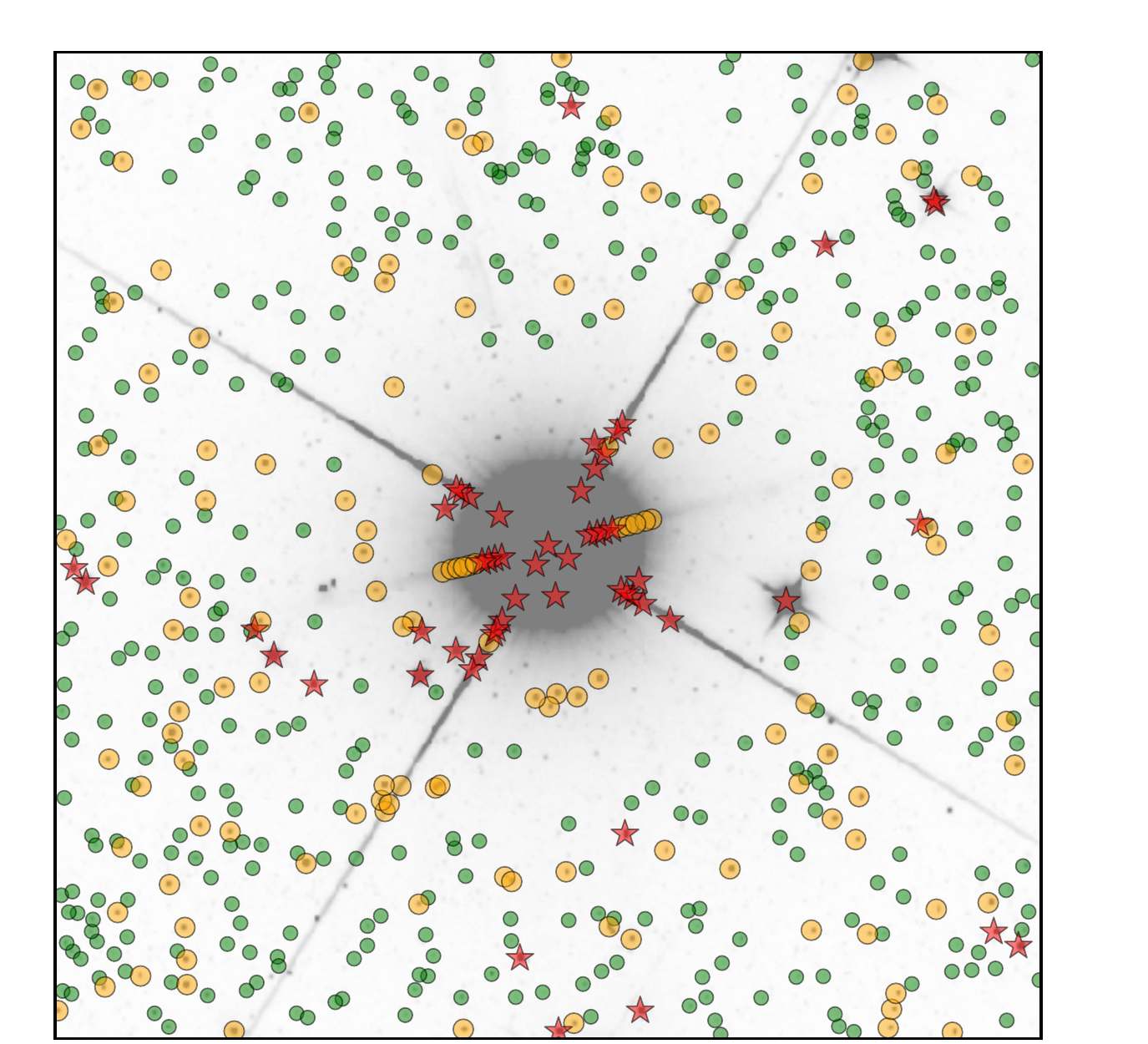} &
      \includegraphics[width=\figw]{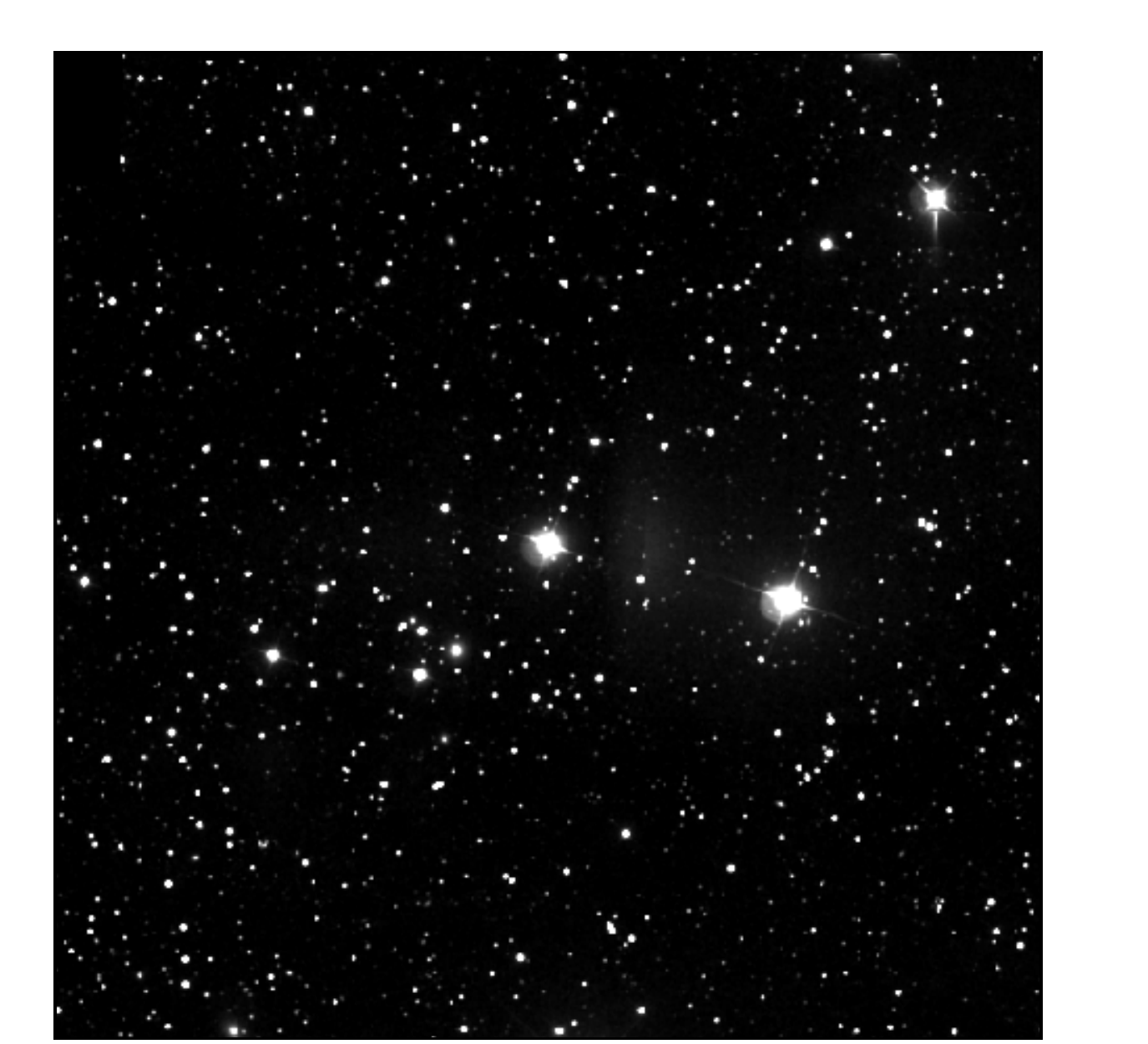} &
      \includegraphics[width=\figw]{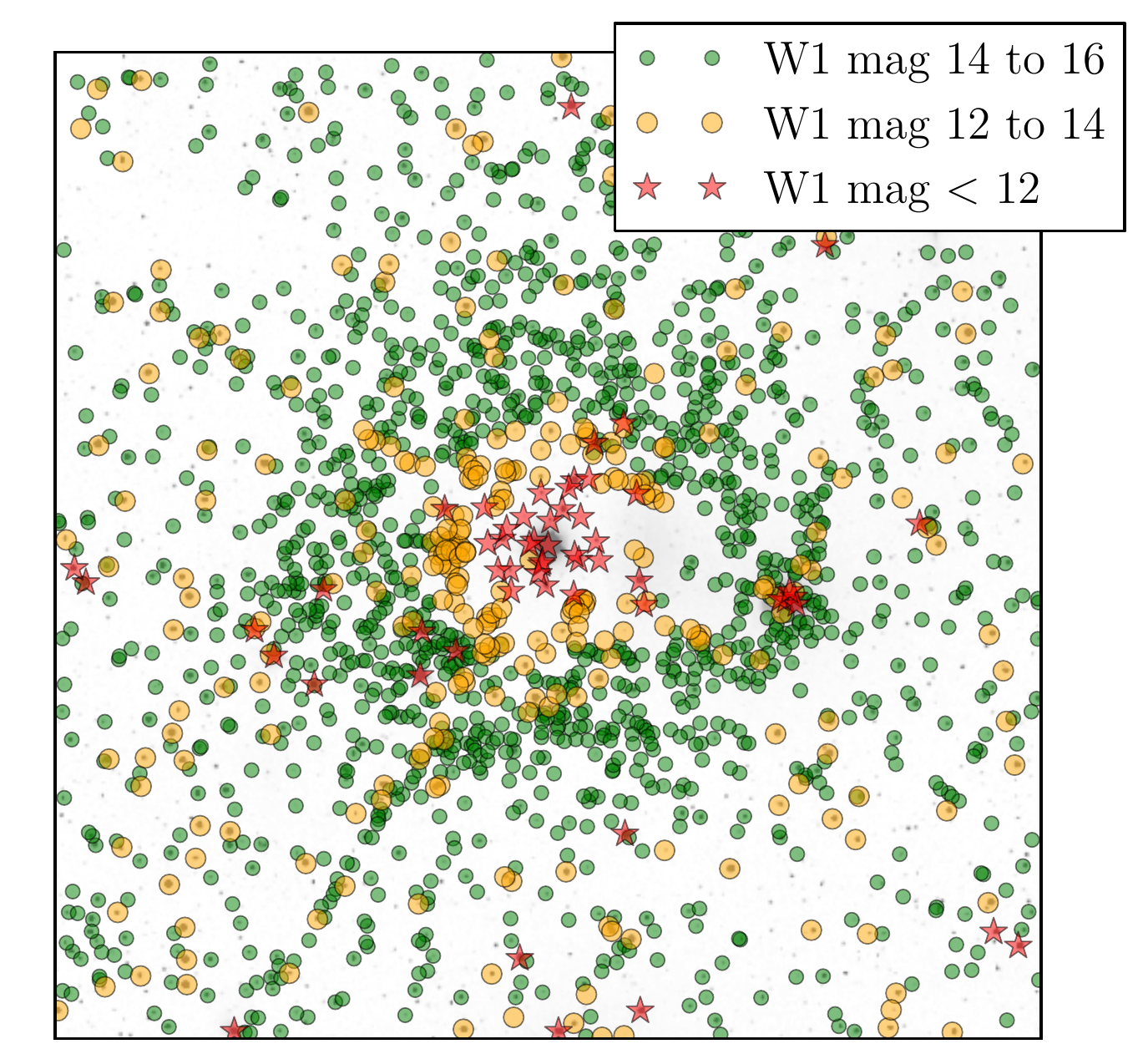} \\
      %
      \includegraphics[width=\figw]{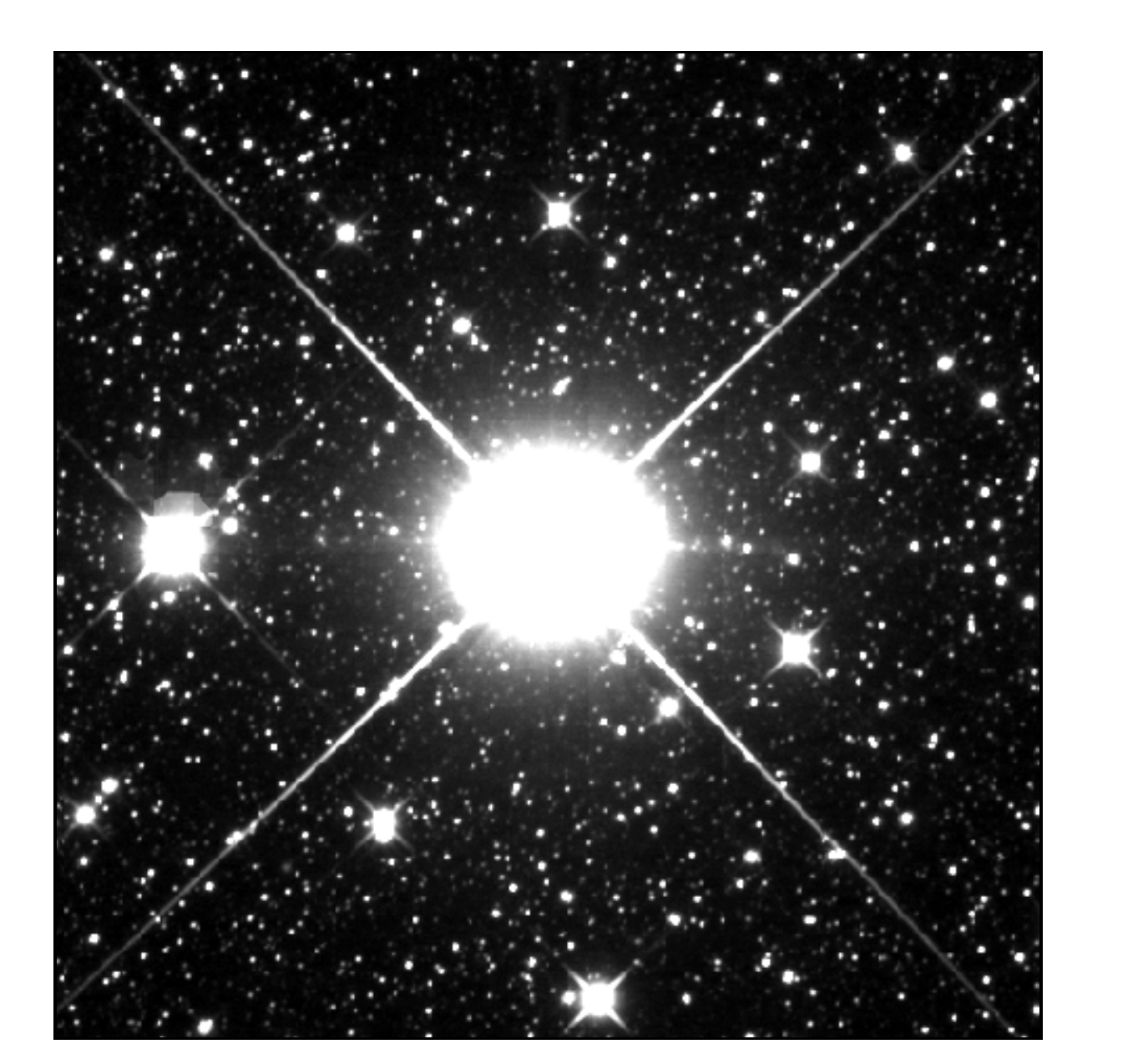} &
      \includegraphics[width=\figw]{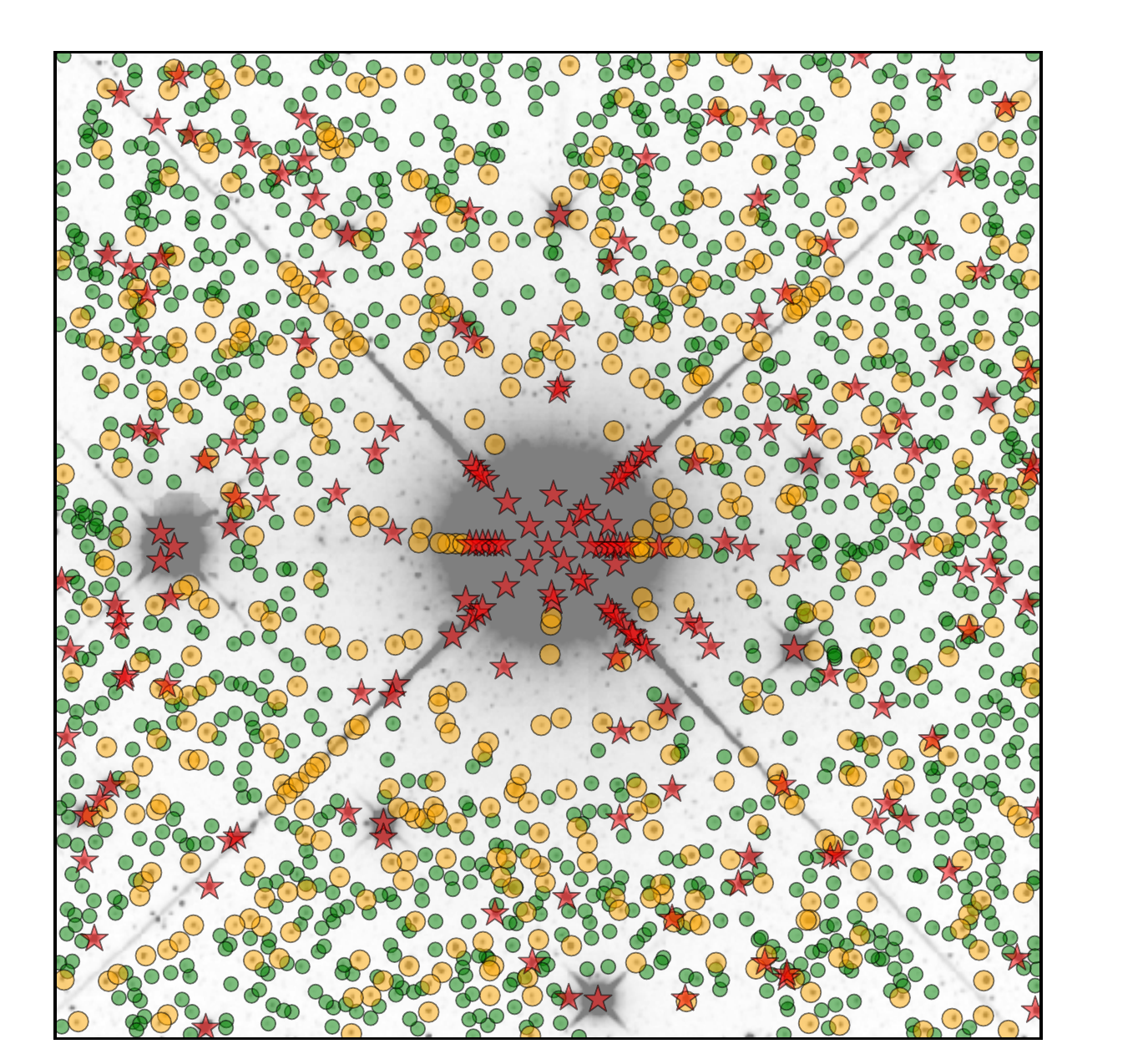} &
      \includegraphics[width=\figw]{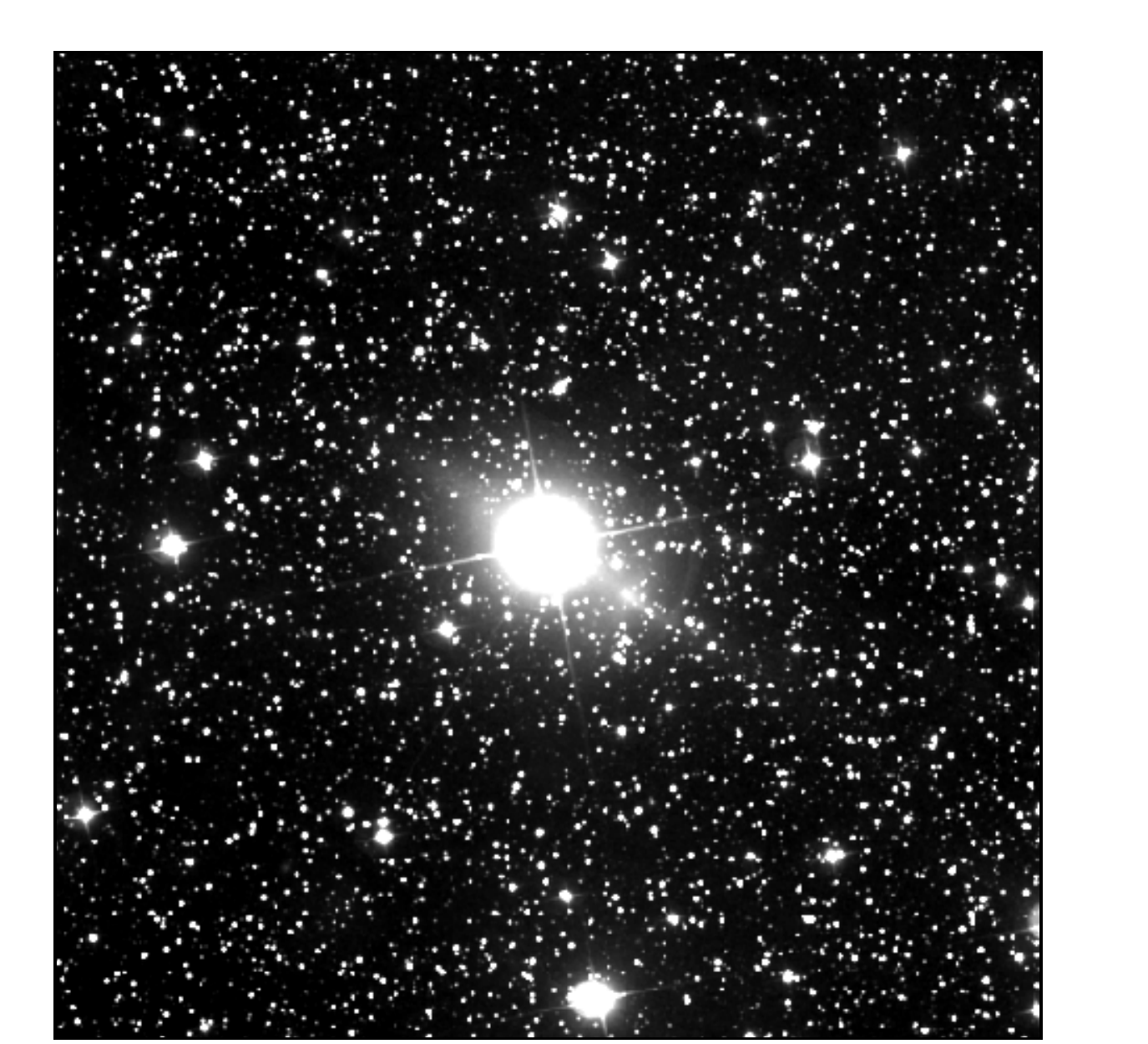} &
      \includegraphics[width=\figw]{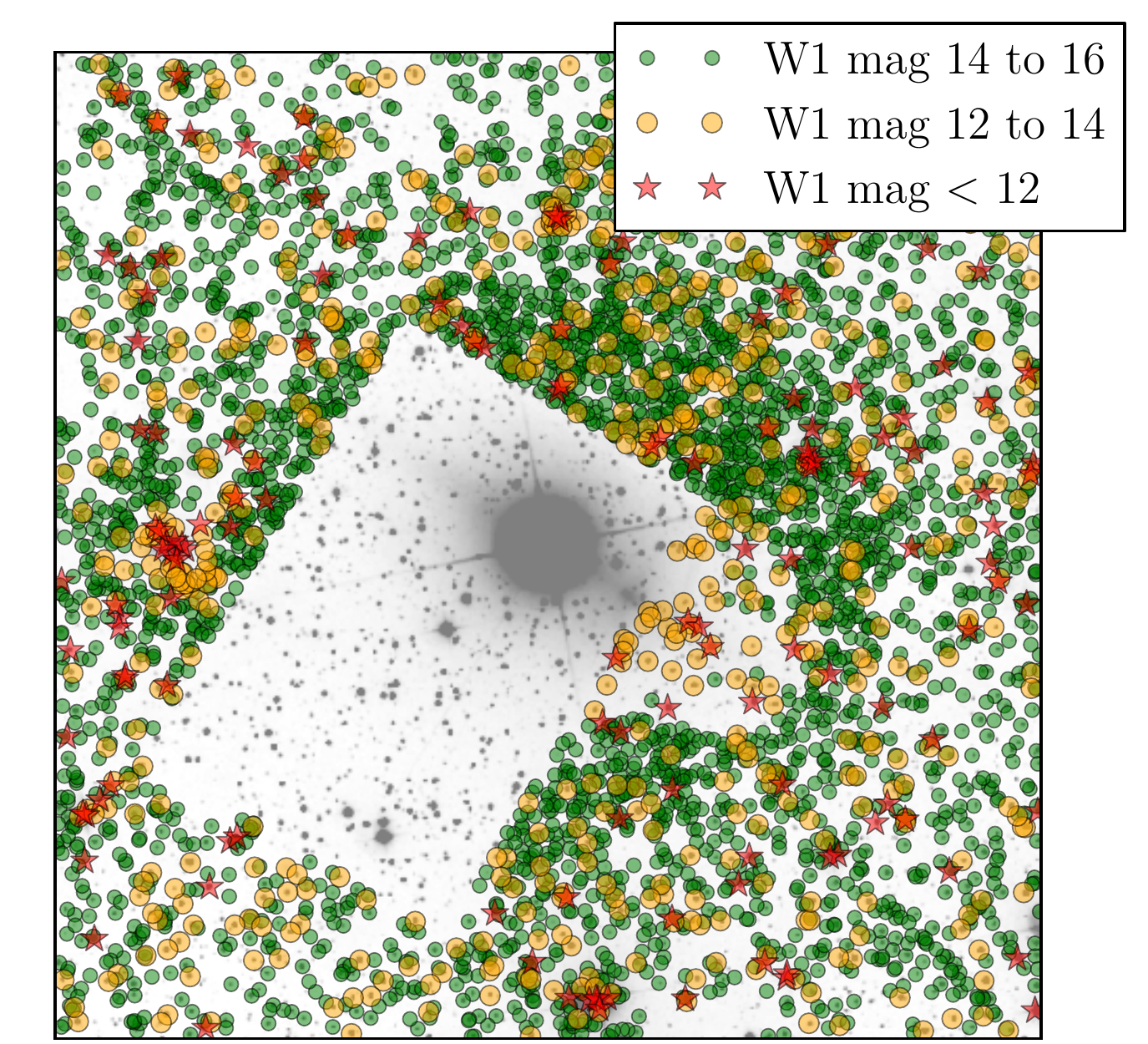} \\
    \end{tabular}
  \end{center}
  \caption{Effects of bright stars on our forced photometry results.
    \textbf{Left column}: WISE W1 images around bright stars.  These
    images are 24 $\times$ 24 arcminutes in size.  \textbf{Second
      column}: AllWISE catalog entries nearby.  Notice the lack of
    moderate-brightness sources near the bright source, and the small
    number of bright artifact sources.  \textbf{Third column}: SDSS
    $r$-band image.  \textbf{Right column}: our forced-photometry
    results.  In the top row, there is a region near the bright star
    that contains few sources in the SDSS catalog (and hence in our
    forced photometry results), a handful of sources measured as being
    very bright, and a ``halo'' of moderately bright sources outside
    the empty region.  In the middle row, the halo of bright
    measurements is very pronounced; the WISE diffraction spikes are
    also apparent.  In the bottom row, the bright star causes the SDSS
    image reduction pipeline to fail, so there are no sources in that
    field.  These examples show extremely bright stars (W1 $\sim -2$);
    fainter stars show more moderate artifacts.
    \label{fig:bright}}
\end{figure}

\Figref{fig:bright} shows a comparison of the artifacts around bright
stars in our forced-photometry results and the AllWISE catalog.  The
most notable effect is a significant halo: sources in the SDSS catalog
that we report as being bright.  This occurs because our PSF models
(as shown in \figref{fig:psf}) focus on the core, largely ignoring the
wings of the PSF.  In our forced photometry approach, when this
unexplained flux coincides with a source, the least-squares fitter
will try to ``explain'' the extra flux by making the source brighter.
As a result, any source near a bright star will appear too bright in our
results.  The examples shown in \figref{fig:bright} are among the
brightest stars in our footprint, and for these stars the halo can be
10 arcminutes or larger.

\section{Discussion}

Forced photometry is one approach for applying information learned in
one survey to data gathered in a second survey.  It is rigid, in the
sense that we \emph{only} photometer the images at locations
containing a source in the input catalog.
As such, forced photometry is most useful when the survey providing
the catalog of sources to photometer has at least the depth and
resolution of the images being photometered.
In addition, forced photometry demands that the images being
photometered are well calibrated.  While it is possible to fit for the
PSF model, astrometric solution and sky level using \thetractor, this
incurs additional computational cost and is not necessarily the most
efficient approach for recalibrating images.  We are fortunate that
the WISE team have produced superbly calibrated images so that image
recalibration has been unnecessary.

A more holistic approach than forced photometry would be to do
\emph{simultaneous fitting}.  We could, for instance, fit all
parameters of the sources (galaxy shapes and positions as well as
fluxes), and include both the SDSS and WISE images in the fitting.
This would extract additional information from both surveys, yielding
stronger constraints on the source properties.  It would also allow
fitting for proper motions and parallaxes of nearby stars.  In
addition, we could detect sources that are below the individual survey
detection thresholds but are significant when the surveys are
combined.  This approach would, however, be significantly more
computationally expensive: To start, SDSS images have roughly 50 times
more pixels per area than WISE.  Further, this would require
non-linear optimization, in contrast to the much cheaper linear
optimization required by forced photometry.  Since we do not expect
the WISE images (with their lower resolution) to have much
constraining power on the positions or shapes of galaxies, an
alternative approach would be first to re-fit the SDSS catalog to the
SDSS images using \thetractor, and then repeat our forced photometry
with that improved catalog.  For moving sources, we could search for
regions of the WISE images that are poorly fit by the SDSS catalog and
allow the sources in these regions to shift their positions slightly.
This post-processing approach would allow us to improve upon the
forced photometry results without increasing the computational cost
excessively.

In this work, we have used coadds of the WISE imaging, rather than the
individual frames.  While in general it would be preferable to
photometer the individual frames (at least in principle), the WISE
images have a stable and approximately isotropic PSF with little
variation over the focal plane, so little information is lost in the
coadding process.  The computational time and memory requirements
scale roughly with the number of pixels being fit, so photometering
the individual frames would have cost roughly 30 times more.

Forced photometry assumes that the profile of a galaxy is the same
between bands.  This does not describe the (typically small) color
gradients across galaxies, although the WISE images lack the
resolution to inform any such gradients in the infrared colors.  Our
forced photometry of the WISE fluxes is equivalent to a
weighted-aperture flux that has the property of being well-defined and
consistently applied to all objects.  Therefore, any biases in the
inferred infrared fluxes would be consistent between galaxies that
have the same intrinsic properties.


In this paper, we used the SDSS $r$-band galaxy shape measurements as
the galaxy profiles for forced photometry.  One might expect the
$z$-band shapes to be closer to the WISE shapes, but the SDSS $z$-band
images are generally of significantly lower signal-to-noise.  Since we
use the same galaxy profiles as used in the $r$-band ``cModelMag''
measurements, our measurements can be used consistently with those
mags.  When the ``fracDev'' deVaucoulers-to-total fraction is zero or
one, our measuments are also consistent with the SDSS ``modelMag''
measurements for all bands.


%
%
%
%
%

\acknowledgements
It is a pleasure to thank Adam Myers, John Moustakas and Abhishek
Prakash for early testing and feedback.

DWH was partially supported by the NSF (grant IIS-1124794), NASA
(grant NNX12AI50G) and the Moore--Sloan Data Science Environment at
NYU.

This publication makes use of data products from the Wide-field
Infrared Survey Explorer, which is a joint project of the University
of California, Los Angeles, and the Jet Propulsion
Laboratory/California Institute of Technology, and NEOWISE, which is a
project of the Jet Propulsion Laboratory/California Institute of
Technology. WISE and NEOWISE are funded by the National Aeronautics
and Space Administration.

This publication makes use of data from the Sloan Digital Sky Survey
III.  Funding for SDSS-III has been provided by the Alfred P. Sloan
Foundation, the Participating Institutions, the National Science
Foundation, and the U.S. Department of Energy Office of Science. The
SDSS-III web site is \niceurl{http://www.sdss3.org/}.

SDSS-III is managed by the Astrophysical Research Consortium for the
Participating Institutions of the SDSS-III Collaboration including the
University of Arizona, the Brazilian Participation Group, Brookhaven
National Laboratory, Carnegie Mellon University, University of
Florida, the French Participation Group, the German Participation
Group, Harvard University, the Instituto de Astrofisica de Canarias,
the Michigan State/Notre Dame/JINA Participation Group, Johns Hopkins
University, Lawrence Berkeley National Laboratory, Max Planck
Institute for Astrophysics, Max Planck Institute for Extraterrestrial
Physics, New Mexico State University, New York University, Ohio State
University, Pennsylvania State University, University of Portsmouth,
Princeton University, the Spanish Participation Group, University of
Tokyo, University of Utah, Vanderbilt University, University of
Virginia, University of Washington, and Yale University.

This research used resources of the National Energy Research
Scientific Computing Center (NERSC), which is supported by the Office
of Science of the U.S. Department of Energy under Contract
No.~DE-AC02-05CH11231.

This research has made use of NASA's Astrophysics Data System.

\appendix

\section{Description of our catalog contents}
\label{sec:catalog}

\newcommand{\filetype}[1]{\textsl{#1}}

Our output files are row-by-row parallel to the SDSS
\filetype{photoObj} files, and are named \filetype{photoWiseForced}.
For example, the SDSS \filetype{photoObj} file containing objects
observed in run 1000, camera column 1, field 100, in data reduction
version 301, is found in the file\footnote{Or at the URL
  \niceurl{http://data.sdss3.org/sas/dr10/boss/photoObj/301/1000/1/photoObj-001000-1-0100.fits}}
\verb+photoObj/301/1000/1/photoObj-001000-1-0100.fits+
and our results are found in the file
\verb+301/1000/1/photoWiseForced-001000-1-0100.fits+
where both of these files contains 368 rows, describing row-by-row the
same objects.

\newcommand{\colname}[1]{\texttt{#1}}

The \filetype{photoWiseForced} files include the following columns:
\parskip0pt
\begin{description}
\topsep0pt
\itemsep1pt  \parsep0pt
\item[\colname{has\_wise\_phot}] (boolean) True for SDSS sources that
  were photometered in WISE.  The object must be PRIMARY in SDSS for
  this to be set.  When this column is False, all other columns have
  value zero.
\item[\colname{ra}, \colname{dec}] (floats) J2000.0 coordinates from
  SDSS.
\item[\colname{objid}] (string) Object identifier from SDSS.
\item[\colname{treated\_as\_pointsource}] (boolean) The SDSS source is
  a galaxy (\colname{objc\_type} == 3) but was treated as a point
  source for the purposes of forced photometry.  If you want an
  optical/WISE color, it would be best to use the SDSS PSF mags, not
  the model mags, for these objects.
\item[\colname{pointsource}] (boolean) The SDSS source is a point
  source (\colname{objc\_type == 6}).
\item[\colname{coadd\_id}] (string) The unWISE coadd tile name, for
  example ``3570p605''
\item[\colname{x}, \colname{y}] (float) Zero-indexed pixel coordinates
  of the source on the unWISE image tile (2048 $\times$ 2048 pixels).
\item[\colname{w1\_nanomaggies}] (float) WISE flux measurement for
  this object.  Note that these are in the native WISE photometric
  system: Vega, not AB.  A source with magnitude 22.5 in the Vega
  system would have a \colname{w1\_nanomaggies} flux of 1.\footnote{
    The WISE team's suggested conversions to AB are available here:
    \niceurl{http://wise2.ipac.caltech.edu/docs/release/allsky/expsup/sec4\_4h.html\#conv2ab}}
\item[\colname{w1\_nanomaggies\_ivar}] (float) WISE formal error as
  inverse-variance.  Note that this formal error does not include
  error due to Poisson variations from the source.  As such, it is
  most appropriate for faint objects.
\item[\colname{w1\_mag}, \colname{w1\_mag\_err}] (floats) Vega
  magnitude and formal error in the forced photometry.  These are
  simple conversions from the ``nanomaggies'' columns above.
\item[\colname{w1\_prochi2}, \colname{w1\_pronpix}] (floats)
  Profile-weighted chi-squared and number-of-pixels values.
  ``Profile-weighted'' means that these are weighted according to the
  profile of the source in the WISE images (eg, weighted by the
  point-spread function for point sources; weighted by the galaxy
  profile convolved by the point-spread function for galaxies).  The
  column \colname{w1\_prochi2} is supposed to measure the quality of
  fit at the location of the source.  Note that \colname{w1\_pronpix}
  effectively counts the fraction of the source that was inside the
  image, and should be close to unity for all sources.
\item[\colname{w1\_proflux}] (float) profile-weighted, the amount of
  flux contributed by other nearby sources.  This will be zero for
  isolated sources, but can be larger that \colname{w1\_nanomaggies}
  if this source is blended with a brighter source.
\item[\colname{w1\_profracflux}] (float) equal to
  \colname{w1\_proflux} divided by \colname{w1\_nanomaggies}; the
  amount of flux at the location of this source that is due to other
  sources, relative to the flux of this source.
\item[\colname{w1\_npix}] (integer) The number of pixels included in
  the fit.
\item[\colname{w1\_pronexp}] (float) The number of WISE exposures
  included in the unWISE coadd at the location of this source.  This
  is a proxy for the depth (which is also reflected in the formal
  error).
\end{description}
plus corresponding columns for \colname{w2}, \colname{w3}, and
\colname{w4}.

\end{document}